\documentclass[a4paper,11pt]{article}
\usepackage[normalem]{ulem}
\usepackage[left=2.5cm,right=2.5cm,top=2cm,bottom=1cm]{geometry}
\usepackage{epsfig,amsfonts,amssymb}
\usepackage{jheppub}
\usepackage{esvect}
\usepackage{comment}
\usepackage{amsmath, amssymb, slashed, epsf, color, graphicx, latexsym, bm}
\usepackage{mathrsfs}
\usepackage{tensor}
\usepackage{physics}
\usepackage{epsfig}
\usepackage{graphics}
\usepackage{mathtools}
\usepackage{caption}
\usepackage{float}
\usepackage{bbm}
\usepackage{subcaption}
\usepackage{enumitem}
\usepackage{verbatim}
\usepackage{blindtext}
\usepackage[mathlines]{lineno}
\usepackage{xcolor}
\usepackage{amsmath}
\usepackage{bm}
\usepackage{framed}
\usepackage{varioref}
\colorlet{shadecolor}{lightgray}
\usepackage{hyperref}
\usepackage{parskip}
\usepackage{soul}

\usepackage[utf8]{inputenc}
\usepackage{soul}

\definecolor{blue}{rgb}{0.0, 0.0, 1.0}

\def \be {\begin{equation}}
\def \ee {\end{equation}}
\def \bea {\begin{eqnarray}}
\def \eea {\end{eqnarray}}

\labelformat{equation}{Eq.~(#1)} 
\usepackage[thinc]{esdiff}
\usepackage{setspace}
\usepackage{tikz}
\usepackage[compat=1.1.0]{tikz-feynman}
\graphicspath{ {./images/} }
\usepackage{xcolor}
\definecolor{PrimaryColor}{RGB}{0, 0, 0} 
\definecolor{Darkblue}{RGB}{0, 0, 128} 
\definecolor{HighlightColor}{RGB}{173, 216, 230} 
\definecolor{AccentColor}{RGB}{128, 0, 0} 
\definecolor{purpleblue}{RGB}{102,59,148} 
\definecolor{Pblue1}{RGB}{105,24,131}
\definecolor{darkteal}{RGB}{1,77,78}
\definecolor{darkgreen}{RGB}{0,168,107}
\definecolor{darkorange}{RGB}{255,83,73}
\begin{document}
\begin{center}

{\Large \bf AdS S-Matrix for Massive Vector Fields}

\end{center}
\vskip .6cm
\medskip

\vspace*{4.0ex}

\baselineskip=18pt

\centerline{\large \rm Nabamita Banerjee$^{a}$, Amogh Neelkanth Desai$^{b}$, Karan Fernandes$^{c,d}$,} 
\centerline{Arpita Mitra$^{e}$, and Tabasum Rahnuma$^{f}$}
\vspace*{4.0ex}
{\small{
\centerline{ \it ~$^a$Indian Institute of Science Education and Research Bhopal,
	Bhopal Bypass, Bhopal 462066, India}
\centerline{\it ~$^b$ Department of Physics, The University of Texas at Dallas, Richardson, Texas 75080, USA. }
\centerline{ \it ~$^c$Department of Physics, National Taiwan Normal University, Taipei, 11677, Taiwan}

\centerline{ \it ~$^d$Center of Astronomy and Gravitation, National Taiwan Normal University, Taipei 11677, Taiwan}

\centerline{ \it ~$^e$ Department  of  Physics,  Pohang  University  of  Science  and  Technology,  Pohang  37673,  Korea}

\centerline{ \it ~$^f$ Asia Pacific Center for Theoretical Physics (APCTP), Pohang, Gyeongbuk 37673, Korea}}}
\centerline{\small Email: nabamita@iiserb.ac.in, Amogh.Desai@utdallas.edu, karanfernandes86@gmail.com, }\centerline{\small arpitamitra89@gmail.com, tabasum.rahnuma@apctp.org}

\vspace*{5.0ex}

\centerline{\bf Abstract} \bigskip

We generalize a recent ``AdS S-matrix" formulation for interacting massive scalars on AdS spacetimes to the case of massive vector fields. This method relies on taking the infinite radius limit for scattering processes perturbatively, which is analyzed using Witten diagrams in the momentum space formulation of global AdS with embedding space coordinates. It recovers the S-matrix with subleading corrections in powers of the inverse AdS radius about a flat spacetime region within the bulk. We first derive the massive vector bulk-to-boundary and bulk-to-bulk propagators within this perturbation theory. As an example, we consider the Abelian Higgs Model in a certain regime of the coupling parameter space to model an interacting Proca theory on AdS spacetimes. We specifically compute the AdS S-matrix for a process involving massive external vector fields mediated by a massive scalar. We lastly discuss possible massless limit of propagators within this perturbative framework.

\baselineskip 18pt

\tableofcontents
\section{Introduction}
The S-matrix is a well-defined observable for quantum field theories in flat spacetime which describes how initial states evolve into final states after scattering processes. Particles in Anti-de Sitter (AdS) spacetime behave like particles in a box. Since massless particles get an effective mass due to the AdS potential, particles with null rays get reflected back from the timelike boundary. Thus in the context of AdS, construction of on-shell asymptotic states are ambiguous and consequently, we lack a well defined notion of an on-shell S-matrix. Scattering phenomena in the bulk of asymptotically AdS spacetimes can be understood in terms of gauge invariant observables which are correlators of a conformal field theory (CFT) at its boundary 
\cite{Maldacena:1997re, Witten:1998qj,Gubser:1998bc, Penedones:2016voo}. There have been several works which discuss how to extract the Infrared (IR) finite S-matrix from large N CFT correlators in the flat space limit of AdS/CFT \cite{Polchinski:1999ry, Giddings:1999jq, Gary:2009mi, Penedones:2010ue, Fitzpatrick:2011ia, Fitzpatrick:2011hu, Fitzpatrick:2011jn, Raju:2012zr, Hijano:2019qmi, Hijano:2020szl, Komatsu:2020sag,   Li:2021snj,Gadde:2022ghy,Marotta:2024sce, Chowdhury:2024wwe, Banados:2024kza}. This has found various applications, including the S-Matrix bootstrap program \cite{Okuda:2010ym, Fitzpatrick:2011hu,Paulos:2016fap, RevModPhys.91.015002,Poland:2016chs,BELAVIN1984333, Maldacena:2015iua}. In AdS, one particle states are irreducible representations of the conformal group, whereas in flat space, they correspond to irreducible representations of the Poincar\'e group. A connection between these two isometry groups in the large AdS radius limit, considered by taking $R \to \infty$, can be obtained by the Inonu-Wigner contraction.  

\par
There exist various frameworks to characterize the flat space limit of boundary CFT correlators within AdS/CFT, such as within momentum space \cite{Raju:2012zr, Gadde:2022ghy}, Mellin space \cite{Penedones:2010ue, Fitzpatrick:2011hu,Fitzpatrick:2011jn, Fitzpatrick:2011ia}, and coordinate space formulations \cite{Okuda:2010ym, Maldacena:2015iua, Komatsu:2020sag}, as well as the partial-wave expansion \cite{Maldacena:2015iua, Paulos:2016fap}. Massive fields in the large $R$ limit have large $\Delta$ (scaling dimension) dual CFT operators, while massless bulk fields in the flat limit are realized from $\Delta \sim d$ boundary operators. The flat limit for massive to massless fields in Mellin space is not smooth due to the presence of an additional saddle point dominating the smearing integral compared to the massless case \cite{Li:2021snj}. In coordinate space, Witten diagrams provide a diagrammatic framework to calculate CFT correlation functions corresponding to bulk scattering processes within AdS spacetime. In general, Witten diagrams comprise of bulk-to-boundary propagators, bulk-to-bulk propagators and interaction vertices. In the flat space limit, the bulk-to-boundary propagator in AdS spacetime reduces to the usual external leg factors for position space Feynman diagrams, while the bulk-to-bulk propagators reduce to Feynman propagators. Consequently, Witten diagrams \cite{Witten:1998qj} also reduce to flat space correlation functions in the large $R$ limit \footnote{In the coordinate space formulation for massless fields, the flat space S-matrix for the contact diagram is obtained from the residue of a 4-point correlator at the ``bulk-point limit'' \cite{Maldacena:2015iua}. On the other hand, the scattering of massive fields in the flat space limit was proposed in \cite{Komatsu:2020sag, Hijano:2019qmi} by relating flat space kinematics to the embedding space coordinates of boundary CFTs.}. These results are valid only in a small central region of global AdS, where distances are very small as compared to AdS radius $R$.
A way to derive a flat space S-matrix in the flat patch in the bulk from position space boundary correlation functions was proposed in~\cite{Komatsu:2020sag} which was proven in~\cite{Li:2021snj} using fixed momentum states constructed from the HKLL prescription \cite{Hamilton:2006az, Hijano:2019qmi}. Subsequently, a perturbative prescription for a S-matrix with subleading corrections in inverse AdS radius in the flat patch was given in \cite{Gadde:2022ghy}, which agrees with~\cite{Komatsu:2020sag} as the leading order result in the large $R$ limit. 
\par
Before discussing this prescription further, we note that one of the  motivations behind studying inverse AdS radius corrections about the flat patch comes from the soft factorization for the S-matrix. The soft limit of a massless gauge field about the flat limit was considered in \cite{Hijano:2020szl}, to relate the soft factorization of the S-matrix \cite{PhysRev.140.B516} in the flat region with the $U(1)$ Ward identity of a boundary CFT. In \cite{Banerjee:2020dww, Banerjee:2021llh, Banerjee:2022oll, Fernandes:2023xim} utilizing a simultaneous double scaling limit, where the frequency of an external leg goes to zero and the AdS radius $R$ approaches infinity while keeping their product finite and large, additional corrections to the leading soft photon and graviton factors were found and later on verified in \cite{Duary:2022pyv, Duary:2022afn} \footnote{Although the AdS radius $R$ serves as an infrared regulator \cite{Fitzpatrick:2011jn} resulting in IR finite observables, treating $1/R$ as a perturbation parameter yields a correction to the leading soft factor.}. These results can be considered as small cosmological constant corrections to soft theorems of flat space scattering amplitudes, which remain less explored compared to those on asymptotically flat spacetimes \cite{Sahoo:2020ryf, Saha:2019tub, PhysRev.135.B1049, PhysRev.140.B516, PhysRev.166.1287, PhysRev.168.1623, He:2014laa,Bern:2014vva, Campiglia:2014yka, Laddha:2017ygw,Klose:2015xoa,  Cachazo:2014fwa, Casali:2014xpa, Banerjee:2022hpo, He:2019jjk}. Soft theorems are also equivalent to Ward identities for asymptotic symmetries, which further constrain physical observables like scattering amplitudes \cite{Strominger:2017zoo,Strominger:2013jfa,He:2014laa,Campiglia:2019wxe, Banerjee:2021uxe, Banerjee:2022lnz}. 
 It would thus be interesting to relate asymptotic symmetries on AdS spacetimes in the large $R$ limit \cite{Compere:2019bua, Compere:2020lrt} with corrections of soft factors about the flat space limit. While this could in principle be explored using Witten diagrams, we note that the prescription in \cite{Gadde:2022ghy} holds exclusively for massive particles. Our analysis in this paper will furnish a generalization for massive vector fields, while leaving the case of massless external fields to future work.

\par
In the present work, our primary focus is to generalize the ``AdS S-matrix'' formulation of \cite{Gadde:2022ghy} to incorporate massive vector fields. 
In \cite{Gadde:2022ghy}, the S-matrix was obtained via the Fourier transform of a position space CFT correlation function in embedding space coordinates, which is defined only for the on-shell momenta. The correlation function is obtained within $1/R$ perturbation theory, where the conformal dimensions of operators dual to the external massive legs grows with $R$ while keeping the mass fixed. Let us first enumerate the necessary steps required to compute the AdS S-matrix for a given massive field theory :
\begin{itemize}
\item We consider AdS${}_{d+1}$ spacetime (with coordinates $X^{\mu}$) which can be embedded in a $(d+2)$ dimensional Minkowski spacetime with metric ${\eta}_{AB}\equiv \rm{diag}(-,+,\dots,+,-)$, 
in a coordinate system $\mathcal{X}^{A}$ constrained by $\mathcal{X}^2 = -R^2$. Note that we denote the coordinates in AdS by Greek indices which runs over 0 to $d$ and Minkowski coordinates by Latin indices which runs over 0 to $d+1$. The AdS boundary, which is a $d$ dimensional space, can also be embedded in the same framework in terms of the embedding space vector ${\mathcal{W}}$ satisfying ${\mathcal{W}}^2=0$.
\item Next we consider a Proca field in the AdS spacetime. Following a boundary Fourier transform defined in terms of conjugate momentum $\mathcal{P}_A$ and imposing a gauge condition on the momentum along $(d+1)-$th direction (${\mathcal{P}}_{d+1} = 0$), we compute both bulk-to-boundary and bulk-to-bulk propagators for the massive vector field perturbatively up to $1/R^2$ order, treating $1/R$ as a perturbative parameter. Note that there exists no appropriate limit where the bulk-to-boundary propagator can be obtained from the bulk-to-bulk propagator at $1/R^2$ order, within this perturbation theory. 
\item As an example, we consider an exchange diagram involving four external massive vector fields with a fixed mass mediated by a massive scalar field, which is realized within an Abelian Higgs model. We compute the AdS S-matrix for this exchange diagram by utilizing the expressions for propagators derived within the $1/R$ perturbation theory.
\item In \cite{Gadde:2022ghy}, it was shown that the momentum space correlation function gets reduced to the S-matrix in the flat space limit, thus proportional to the momentum-conserving delta function, $\delta(\sum P_i)$, at leading order in $1/R$. As translation symmetry is broken beyond the flat space limit due to the AdS potential, the correlation function also contain terms proportional to derivatives of $\delta(\sum P_i)$ at subleading orders in $1/R$.
In particular, the AdS S-Matrix is defined as the term which is proportional to $\delta(\sum P_i)$.
 
\end{itemize}
 \par
The relevant expressions are well defined in a local patch on the AdS hyperboloid, which is centered around a specific local point $C=(0,\dots,0, R)$ in the large $R$ limit.  This is a consequence of the ${\mathcal{P}}_{d+1} = 0$ choice mentioned above and fixes a conformal frame for the momentum space correlator \cite{Gadde:2022ghy}. Once we fix the gauge for the AdS momenta variables, all the bulk processes are confined and localized within this frame surrounding the point $C$, where we represent the AdS momenta variables as the flat space momenta, and the isometry algebra can be shown to be that of flat spacetime. More specifically, this is realized through the contraction of the conformal generators $M_{\mu, d+1} \to R P_{\mu}$. An important feature of the AdS S-matrix as defined is that it can recover the corresponding momentum space boundary correlation function. This is achieved through the inverse AdS radius corrections about the flat patch, which through the Ward identity for $M_{\mu,d+1}$, recover terms involving derivatives of the momentum conserving delta function.  

\par
The paper is organized as follows.
Section \ref{Bbdyprop} contains the construction of bulk-to-boundary and bulk-to-bulk propagator for massive vector fields within $1/R$ perturbation theory.  
We have computed both the propagators up to subleading order in perturbation around the flat space. 
In section \ref{brief}, we compute one AdS S-matrix for an Abelian Higgs model in AdS$_{d+1}$. Then in section \ref{limits}, we discuss possible massless and double scaling limits for the propagators, which is motivated by analogous limits considered for soft limits in AdS using momentum space formalisms.
Lastly, in section \ref{conc}, we conclude our results with some future directions. Two appendices detail the derivation of the bulk-to-boundary propagator in $1/R$ perturbation theory (Appendix \ref{appen1}) and intermediate steps in the AdS S-matrix derivation for the scattering of four massive vector fields mediated by a massive scalar (Appendix \ref{appen2}). 

\section{Massive gauge fields and their propagators on AdS$_{d+1}$}\label{Bbdyprop} 

In this section, we will consider a massive vector field on a $(d+1)$ dimensional AdS spacetime and derive its bulk-to-boundary and bulk-to-bulk propagators within $1/R$ perturbation theory around a centrally located flat patch in global coordinates. The AdS$_{d+1}$ spacetime can be described by embedding it in a $(d+2)$ dimensional Minkowski spacetime $\eta_{AB} = \text{diag} (-, + , \cdots, -)$. We denote the coordinates of $AdS_{d+1}$ of length $R$ by a set of points $X \equiv(X^{0}, X^{1}, \cdots, X^{d})$ and the spacetime interval for the Minkowski spacetime is given by,
\begin{equation}
d s^2=-\left(d X^0\right)^2+\left(d X^1\right)^2+\cdots+\left(d X^{d}\right)^2-\left(d X^{d+1}\right)^2
\label{emb.met}
\end{equation}
with coordinates $\mathcal{X}^A=(X, X^{d+1})$. In embedding coordinates, $AdS_{d+1}$ is a hyperboloid satisfying $\mathcal{X}^A\mathcal{X}_A=-R^2$ with its boundary described by the projective null cone $\mathcal{W}^A\mathcal{W}_A= 0$. To distinguish the AdS$_{d+1}$ spacetime metric, we define $W=X^{d+1}$ such that \ref{emb.met} takes the form  
\begin{equation}
d s^2=\eta_{\mu\nu} d X^{\mu} d X^{\nu}-d W^2 
\end{equation}
The hyperboloid constraint condition further implies that $W^2=R^2+X^2$, with $X^2=X_{\mu}X^{\mu}$. Since we are interested treating $1/R$  as a perturbative parameter and studying corrections up to $1/R^2$ order, we will consider the metric and its inverse as,
\begin{equation}
 g_{\mu\nu}=\eta_{\mu\nu}-\frac{X_{\mu} X_{\nu}}{R^2+X^2}, \quad g^{\mu\nu}(X)=\eta^{\mu\nu}+\frac{X^\mu X^\nu}{R^2} 
 \label{met}
\end{equation}
The covariant derivatives and affine connection are given by
 \begin{equation}
\begin{aligned}
\nabla_{X} & =\partial_{X}-\Gamma, \quad \Gamma_{\alpha \beta}^{\gamma}=-\frac{1}{\mathrm{R}^2} X^{\gamma} g_{\alpha\beta}.
\label{cov}
\end{aligned}
 \end{equation}
We note that the connection in \ref{cov} and inverse metric in \ref{met} are exact to all orders in $1/R$.
We will now consider a massive vector field $A^{\mu}$ with mass $M$ in $\text{AdS}_{d+1}$, which is described by the Proca action on the AdS background \cite{Kabat:2012hp, Costa:2014kfa},
\begin{align}\label{ActionP}
    S_{M}=-\int d^{d+1}X\sqrt{-g}\left[\frac{1}{4} F_{\mu\nu}F^{\mu\nu}+\frac{1}{2}M^2 A_{\mu}A^{\mu}\right] \,,
\end{align}
where $F_{\mu\nu}=\nabla_{\mu}A_{\nu}-\nabla_{\nu}A_{\mu}$. The action provides the following equation of motion for $A^{\mu}$,

\begin{align}\label{PE}
    \nabla_{\mu}F^{\mu\nu}-M^2 A^{\nu}=0\,,
\end{align}
which on using \ref{met} and \ref{cov} gives

\begin{equation}\label{eqm}
   \left[g_{\mu \nu}(\nabla^2_X- M^2) - \nabla_{\nu}\nabla_{\mu} \right]A^{\nu} =0.
\end{equation}
This equation of motion will be used to derive the bulk-to-bulk and bulk-to-boundary propagators for the massive vector field. Lastly, we note that the Proca field by virtue of \ref{PE} further satisfies the condition

\begin{equation}
\nabla_{\mu}A^{\mu}=0.
\label{lc}
\end{equation}

It is known that the massless limit of \ref{ActionP} will require addressing an additional longitudinal degree of freedom, which can be achieved through the the St\"ukelberg trick \cite{Stueckelberg:1957zz,Hinterbichler:2011tt}. However, there are more fundamental obstructions towards taking a massless limit within the AdS S-matrix formalism, as noted previously in \cite{Gadde:2022ghy}. We will make further comments on taking this limit in section \ref{limits}.\\

\subsection{Bulk-to-Boundary Propagator}\label{Bbdyprop1} 
In general for AdS spacetimes in the embedding formalism, the bulk-to-boundary propagator, denoted by $E^{\nu\beta}_{\Delta}(\mathcal{X},\mathcal{W})$ between two spinning fields inserted at a local bulk point $\mathcal{X}$ and a point on boundary $\mathcal{W}$ is proportional to  $\left(\mathcal{X}\cdot\mathcal{W}\right)^{-(\Delta+J)}$ for spin $J$ \cite{Allen:1985wd, Naqvi:1999va,Costa:2014kfa, Costa:2011mg}. In the AdS S-matrix formulation using $1/R$ perturbation theory, the bulk-to-boundary propagator must incorporate boundary momenta. This ensures that the external leg momenta in the bulk S-matrix remain on-shell while taking the flat limit. This is further realized from a momentum space expression which can be derived by performing an appropriate Fourier transform on the boundary coordinate, i.e.

\begin{equation}
\widetilde{\Pi}^{\nu \beta}\left(\mathcal{P}\,,\mathcal{X}\right) = \int d^{d+2} \mathcal{W}~ \delta (\mathcal{W}^2) e^{i \mathcal{P}\cdot \mathcal{W}} E^{\nu\beta}_{\Delta}(\mathcal{X},\mathcal{W}) 
\label{bbp}
\end{equation}

We now recall some properties of \ref{bbp} that provide external states in the AdS S-matrix with fixed momenta. These are similar to properties noted in \cite{Gadde:2022ghy} for the bulk-to-boundary propagator for massive scalar fields. The property $E^{\nu\beta}_{\Delta}(\mathcal{X}, \lambda \mathcal{W}) = \lambda^{-\Delta} E^{\nu\beta}_{\Delta}(\mathcal{X}, \mathcal{W})$ for $\lambda>0$, follows from the homogeneity of embedding space fields \cite{Costa:2011mg}, and hence we have

\begin{align}
\tilde{\Pi}^{\nu \beta}\left( \lambda \mathcal{P}\,,\mathcal{X}\right) = \lambda^{\Delta - d} \tilde{\Pi}^{\nu \beta}\left( \mathcal{P}\,,\mathcal{X}\right) \Rightarrow \tilde{\Pi}^{\nu \beta}\left( \hat{\mathcal{P}}\,,\mathcal{X}\right) = \vert \mathcal{P} \vert^{d- \Delta} \tilde{\Pi}^{\nu \beta}\left( \mathcal{P}\,,\mathcal{X}\right)
\label{orm}
\end{align}

The transformation in \ref{orm} ensures that we can derive vector bulk-to-boundary propagators as a function of the orientation $\hat{\mathcal{P}} = \frac{\mathcal{P}}{\vert \mathcal{P} \vert}$ on setting $\lambda = \frac{1}{\vert \mathcal{P} \vert}$, where $\vert \mathcal{P} \vert$ is the magnitude. 
In the case of massive fields setting $\mathcal{P}_{d+1} = 0$ yields $\mathcal{P}^2 = -M^2 = P^2$. In addition, it imposes the embedding position to take the form of $\mathcal{X} = \left(X\,, \sqrt{X^2 + R^2}\right)$ and this provides a perturbative $1/R$ expansion about flat spacetime in the $R \to \infty$ limit. Note that this property is  also consistent with the leading contribution of \ref{bbp} in the large $R$ limit as noted in \cite{Komatsu:2020sag, Gadde:2022ghy}. On rescaling $\mathcal{W} \to R \mathcal{W}$ and setting $\mathcal{W} = (W , 1)$ in \ref{bbp} we get the following dominant contribution in the large $R$ limit,
\begin{align}
\widetilde{\Pi}^{\nu \beta}\left(P\,,X\right) \sim& \quad\frac{R^{\Delta - d+1}}{2 \pi} \int d^{d+1}{W} \int d \beta~\, e^{i R \beta (W^2 -1)} e^{i R P\cdot W} \alpha^{\nu \beta}\left(P\,,X\right) \notag\\
& \quad \left( W\cdot X - \sqrt{X^2 + R^2}\right)^{-M R}\,,
\label{bbp2}
\end{align}
with a $\beta$ integral representation for the Dirac delta function. In \ref{bbp2} we have utilized that $E^{\nu\beta}_{\Delta}(\mathcal{X},\mathcal{W}) \sim\left(\mathcal{X}\cdot\mathcal{W}\right)^{-(\Delta+J)}$, and we made use of the large $\Delta = M R$ limit to approximate $$\lim_{\Delta \gg 1} E^{\nu\beta}_{\Delta}(\mathcal{X},\mathcal{W}) \approx \alpha^{\nu \beta}\left(P\,,X\right) \left( W\cdot X - \sqrt{X^2 + R^2}\right)^{-M R}.$$ The integrand contribution from the second line of \ref{bbp2} can be further approximated by
\begin{equation}
\left( W\cdot X - \sqrt{X^2 + R^2}\right)^{-M R} \approx (-R)^{-M R} e^{M W\cdot X}
\label{intbb}
\end{equation}
where notably the exponential term does not involve $R$.
On the other hand, the first line of \ref{bbp2} does involve exponential terms that provide the saddle point contribution $W^{\mu}_* = i \frac{P^{\mu}}{M}$ in the large $R$ limit \footnote{The saddle is $2\beta W_{*}^{\mu}=-P^{\mu}$. We can fix $2\beta=iM$ from $W^2=1$.}. Using this saddle contribution in \ref{bbp2} along with \ref{intbb}, we get $\widetilde{\Pi}^{\nu \beta}\left(P\,,X\right)  \propto \alpha^{\nu \beta}\left(P\,,X\right) e^{i P.X}\,$. 
Hence in the large $R$ limit, the leading contribution from the Fourier transformed bulk-to-boundary propagator are plane wave states. As a consequence, the bulk-to-boundary propagator with the boundary point Fourier transformed to ``momentum" space in embedding coordinates gives fixed bulk on-shell momenta states in the large $R$ limit. However, the derivation of perturbative corrections in $1/R$ from the Fourier transform, which will come from expanding $\mathcal{X}$, is more involved using the saddle point approximation analysis. A simpler alternative towards perturbative $1/R$ corrections in the bulk-to-boundary propagator comes directly from solving the source-free equations of motion of the field in AdS$_{d+1}$. 

We solve the propagator in the large $R$~ limit with an iterative treatment to  derive $1/R$ perturbations about the leading flat space limit. 
\ref{eqm} implies that the bulk-to-boundary propagator $\widetilde{\Pi}^{\nu \beta}(P; X)$ satisfies the following differential equation,
\begin{equation} \label{prop1}
    \left[ g_{\mu \nu} \left(\nabla^2_X - M^2\right) - \nabla_\nu \nabla_\mu \right]  \widetilde{\Pi}^{\nu \beta}(P; X)= 0.
\end{equation}

\par

where $\nabla_X$ was defined in \ref{cov}. After a bit more simplification we obtain
\begin{align}\label{eqw}
& [\eta_{\mu\nu}(\partial_X^2-M^2)-\partial_\nu \partial_\mu ]\widetilde{\Pi}^{\nu \beta}\notag\\&+{1\over R^2} \Big[ -X_{\mu}X_{\nu}\Big(\partial_X^2 -M^2\Big)\widetilde{\Pi}^{\nu \beta}+\Big(\eta_{\mu\nu}X^{\rho}X^{\sigma}\partial_{\rho}\partial_{\sigma}-2 \eta_{\mu\nu} -2 X_{\mu}\partial_{\nu}+X_{\nu}\partial_{\mu}\Big)\widetilde{\Pi}^{\nu\beta}\notag\\&+\Big((d+1)X.\partial_X+(d+2)\Big)\eta_{\mu\nu}\widetilde{\Pi}^{\nu\beta} +X^{\beta}\eta_{\rho\mu}\partial_{\nu}\widetilde{\Pi}^{\nu\rho}-2\eta_{\mu\nu}X^{\beta}\partial_{\alpha}\widetilde{\Pi}^{\nu\alpha}+X^{\beta}\partial_{\mu}\widetilde{\Pi}\Big]=0
\end{align}
where we have used the following expansion of the Laplacian
\[
\begin{split}
     \nabla^2_X \widetilde{\Pi}^{\nu \beta}=g^{\rho\sigma}\nabla_{\rho} \nabla_\sigma \widetilde{\Pi}^{\nu \beta}=
       \partial^2 \widetilde{\Pi}^{\nu \beta}-{1\over R^2}&\Big[-X^{\rho}X^{\sigma}\partial_{\rho}\partial_{\sigma}\widetilde{\Pi}^{\nu \beta}+2  \widetilde{\Pi}^{\nu \beta}+2 X^{\nu}\partial_{\alpha}\widetilde{\Pi}^{\alpha\beta}\notag\\&+2 X^{\beta}\partial_{\alpha}\widetilde{\Pi}^{\nu\alpha}-(d+1)X^{\alpha}\partial_{\alpha}\widetilde{\Pi}^{\nu\beta}\Big].
\end{split}
\]
and the last second order derivative operator acting as
\begin{align}
\nabla_{\nu} (\nabla_\mu \widetilde{\Pi}^{\nu \beta})=
\partial_\nu \partial_\mu \widetilde{\Pi}^{\nu \beta} - \frac{1}{R^2}\Big[\Big((d+2) \eta_{\mu \nu} + {X_\nu \partial_\mu\Big)\widetilde{\Pi}^{\nu \beta} }+{X^{\beta}\partial_{\mu}\widetilde{\Pi}}+{X^{\beta}\eta_{\rho\mu}\partial_{\nu}\widetilde{\Pi}^{\nu\rho}}\Big] 
\end{align}
for the coordinate system chosen in \ref{met}. It will be convenient to consider the bulk-to-boundary solution in terms of the momentum direction by defining $\eta=i\hat{P}\cdot X$, with the unit momentum vector $\hat{P}= \frac{P}{|P|}$. 
  
One can assume that in a general scattering process, the perturbative solution of \ref{eqw} takes the form

\begin{equation}\label{fullbdy}
    \widetilde{\Pi}^{\nu \beta} (\hat{P} \,, X) =e^{iM\eta}\left[G^{\nu \beta}(\hat{P}) + \frac{1}{R^2}H^{\nu \beta} (\hat{P} \,,X) + \mathcal{O} \left(R^{-3}\right)\right]. 
\end{equation}

It is evident from \ref{eqw} that the propagator has two solutions depending on the choice $P=\pm iM \hat{P}$. On physical grounds, we discard the negative energy solution from the onset. Therefore we set $e^{iM\eta}=e^{iP\cdot X}$. We have provided the detailed derivations of the leading and subleading contributions of \ref{fullbdy} in Appendix \ref{appen1}. In the following, we summarize the main steps and quote the results.

\par
Substituting \ref{fullbdy} in \ref{eqw}, we obtain the leading order equation,
\begin{equation}\label{diffff}
    \left[  \eta_{\mu \nu}(\partial^2_X - M^2)- \partial_\mu  \partial_\nu \right]e^{iM\eta}G^{\nu \beta}(\hat{P})=0.
\end{equation}

\ref{diffff} can be readily solved to get
\begin{equation}
G^{\nu \beta}(\hat{P})=\eta^{\nu\beta}-\hat{P}^{\nu}\hat{P}^{\beta},
\label{G.def}
\end{equation}
which satisfy the following properties
\begin{equation}
 \eta_{\mu \nu}G^{\mu \nu}(\hat{P})=d, \qquad  {P}_{\nu}  G^{\nu \beta}(\hat{P})=0.
\end{equation}

To next order, we have an equation with a $1/R^2$ correction, which follows from the perturbative expansion of \ref{eqw} 
\begin{align}\label{subleadingsec1}
& \eta_{\mu\nu}\Big(-2 M(\hat{P}\cdot\partial)+\partial^2_X\Big) H^{\nu\beta}(P\,,X)-\left(\partial_{\nu} \partial_{\mu}+M^{2} \hat{P}_{\mu} \hat{P}_{\nu}-M\left(\hat{P}_{\mu} \partial_{\nu}+\hat{P}_{\nu} \partial_{\mu}\right)\right)H^{\nu\beta}(P\,,X)\notag\\
& +\left(d-(d+1)M(X\cdot\hat{P})+M^{2}(X\cdot\hat{P})^{2}\right)\eta_{\mu\nu} G^{\nu\beta}(P)-M X^{\beta} \hat{P}_{\mu}(d+1)+M(X\cdot \hat{P}) \hat{P}^{\beta} \hat{P}_{\mu}=0
\end{align}
We can solve \ref{subleadingsec1} by assuming a general ansatz for $H^{\nu \beta}$ as a linear combination of relevant two-index tensor quantities, i.e. $\eta^{\nu \beta}, P^\nu P^\beta, P^\nu X^\beta, X^\nu P^\beta$. The derivation has been provided in Appendix \ref{appen1} and we find the solution 

\begin{equation}\label{VectBulkToBdry}
    \begin{split}
        H^{\nu \beta}(\hat{P}\,,X) &= \left[-\frac{(d+2) (\hat{P}\cdot X)}{4M} -\frac{d (\hat{P}\cdot X)^2}{4} + \frac{M (\hat{P}\cdot X)^3 }{6}\right] G^{\nu \beta} (\hat{P}) \\
&\qquad \quad  - \frac{(d+1)}{M } X^{\beta} \hat{P}^{\nu} + \frac{1}{M} X^{\nu} \hat{P}^{\beta} + \frac{d}{M^2} \hat{P}^{\beta} \hat{P}^{\nu}.
    \end{split}
\end{equation}
Hence, the bulk-to-boundary propagator as a function of $({P},X)$ up to the first subleading correction in $1/R^{2}$ takes the form
\begin{equation}\label{fullbdyP}
    \begin{split}
         \widetilde{\Pi}^{\nu \beta}(\hat{P},X)=&e^{i P\cdot X}\left[G^{\nu \beta}(\hat{P})+ {1\over R^2} H^{\nu \beta}(\hat{P}\,,X)\right]\\ =&e^{i P\cdot X}\left[ G^{\nu \beta}(\hat{P}) + \frac{1}{R^2} \Bigg\{\left( -\frac{(d+2) (\hat{P}\cdot X)}{4M} - \frac{d (\hat{P}\cdot X)^2}{4} + \frac{M (\hat{P}\cdot X)^3}{6} \right) G^{\nu \beta}(\hat{P}) \right. \\
         & \left. \qquad \quad - \frac{(d+1)}{M} X^{\beta} \hat{P}^{\nu} + \frac{1}{M} X^{\nu} \hat{P}^{\beta} + \frac{d}{M^2} \hat{P}^{\beta} \hat{P}^{\nu} \Bigg\} \right]  
    \end{split}
\end{equation}

 In the following subsection, we will first derive the bulk-to-bulk propagator for massive vector fields. We will then use these results in the next section to compute the Witten diagram for a scattering process.

\par

\subsection{Bulk-to-Bulk Propagator}\label{BBpro}

This propagator is a two-point correlation function of a given field theory, interpreted physically as its propagation between any two points $X$ and $Y$ in the bulk of AdS. We are interested in the propagators for massive vector and scalar fields, which are defined using the time-ordered product (T-product) of two operators as below, 

\vspace{2em}
\begin{figure}[hbt]
\begin{minipage}{0.3\textwidth}
\begin{tikzpicture}
    \begin{feynman}
    \diagram[horizontal=a to b] {
  a -- [scalar, very thick, Darkblue] b [xshift=1.5cm] };
    
         \node at (a) [below=0.1] {$X_1$};
        \node at (b) [below=0.1]{$X_2$};
\end{feynman}
\end{tikzpicture}
    \end{minipage}
    \begin{minipage}{0.3\textwidth}
    \vspace{-5ex}   
  \begin{equation*}
 \hspace{2ex}: \qquad \quad  G(X_1, X_2) \equiv \langle T \{\Phi(X_1)\Phi(X_2)\} \rangle.
  \end{equation*}
\end{minipage}
\end{figure}
    \vspace{-4ex}   
\begin{figure}[hbt]
\begin{minipage}{0.3\textwidth}
\begin{tikzpicture}
    \begin{feynman}
    \diagram[horizontal=a to b] {
 
  a -- [boson, very thick, Darkblue] b [xshift=1.5cm]};
     \node at (a) [above=0.1] {$\mu$};
        \node at (b) [above=0.1]{$\nu$};
         \node at (a) [below=0.1] {$X_1$};
        \node at (b) [below=0.1]{$X_2$};
\end{feynman}
\end{tikzpicture}
    \end{minipage}
    \begin{minipage}{0.3\textwidth}
    \vspace{-2ex}   
  \begin{equation*}
\hspace{2ex} : \qquad  \Pi^{\mu\nu}(X_1, X_2) \equiv \langle T \{ A^{\mu}(X_1)A^{\nu}(X_2)\} \rangle.
  \end{equation*}
\end{minipage}

\caption{Definition of vector and scalar propagators in Bulk AdS.}
\label{defprops}
\end{figure}

The momentum space bulk-to-bulk vector propagator, with one of the positions Fourier transformed to momentum space, will be important in our consideration of Witten diagrams. The bulk-to-bulk propagator $g_{BB}(\tilde{P}, X)$ for a scalar field in AdS involving $R^{-2}$ corrections was derived  in \cite{Gadde:2022ghy}. While $\tilde{P}$ is defined as the momentum about a locally flat region within AdS in the large $R$ limit, it is not on-shell as in the the bulk-to-boundary propagator case. To derive $G_{BB}^{\mu \nu}(\tilde{P}, X)$ for the massive vector field, we work with the equations of motion satisfied by the position space bulk-to-bulk propagator
 \begin{equation}
 \Big[ g_{\mu \nu}(\nabla_{X}^2 - M^2)- \nabla_{\nu } \nabla_{\mu}\Big]G_{BB}^{\nu \beta}(Y,X) = -\frac{1}{\sqrt{g (X)}}\delta_{\mu}^\beta \delta(Y\,,X) \,,
 \end{equation}
and Fourier transform $Y$ to momentum space to find 
 
\begin{equation}\label{eombb}
     \Big[ g_{\mu \nu}(\nabla_{X}^2 - M^2)- \nabla_{\nu } \nabla_{\mu}\Big]G_{BB}^{\nu \beta}(\tilde{P},X) = -\frac{1}{\sqrt{g(X)}}\delta_{\mu}^\beta e^{i \tilde{P} \cdot X} 
\end{equation}

\par

As in the derivation of the bulk-to-boundary propagator, we consider $G^{\nu \beta}_{BB}$ perturbatively expanded in powers of $1/R^{2}$ such that
\begin{equation}\label{BBprop}
    G^{\nu \beta}_{BB} = e^{i\tilde{P} \cdot X}\left[ G^{\nu \beta}_{BB(1)} + \frac{1}{R^2} G^{\nu \beta}_{BB(2)}+ \dots \right]
\end{equation}
The leading contribution can be derived using the ansatz 
\[G_{BB(1)}^{\nu \beta}(\tilde{P},X)=(A(\tilde{P},X) \eta^{\nu\beta}+B(\tilde{P},X) \tilde{P}^{\nu}\tilde{P}^{\beta}) \;e^{i\tilde{P}.X}.\]
On substituting this in the leading term of \ref{eombb}, we find the solution
\begin{equation}\label{f1}
G_{BB(1)}^{\nu \beta}(\tilde{P},X) = \left(\eta^{\nu \beta} + \frac{\tilde{P}^\nu \tilde{P}^\beta}{M^2}\right)\frac{e^{i \tilde{P} \cdot X}}{\tilde{P}^2 + M^2} \,,
\end{equation}
which is the leading contribution of the bulk-to-bulk massive vector propagator in a locally flat region of the AdS spacetime.

To solve the equation to subleading $R^{-2}$ order, we  use equation \ref{BBprop} in \ref{eombb}, which provides
\begin{equation}\label{eomoffshell}
\begin{split}
&\Big[\eta_{\mu \nu}(\partial^2 -M^2)- \partial_{\mu} \partial_{\nu}\Big]e^{i \tilde{P} \cdot X}G_{BB(2)}^{\nu \beta}- 2 \eta_{\mu \nu} X^{\nu}( \partial_{\alpha} e^{i\tilde{P} \cdot X})G_{BB(1)}^{\alpha \beta }\\
&+\Big[ -X_{\mu}X_{\nu}(\partial^2 - M^2 )+ \eta_{\mu \nu}\Big( d+ (X \cdot \partial_X )^2 + d X \cdot \partial_X \Big) + X_{\nu } \partial_{\mu} 
\Big]e^{i \tilde{P} \cdot X}G_{BB(1)}^{\nu \beta}\\ & - X^{\beta} (\partial_{\alpha} e^{i\tilde{P} \cdot X})G_{BB(1)}^{\alpha \nu } + \eta_{\rho \sigma}X^\beta (\partial_{\mu} e^{i \tilde{P} \cdot X} )G_{BB(1)}^{\rho \sigma} = -\delta_{\mu}^\beta \frac{X \cdot X}{2}e^{i\tilde{P}.X}.
\end{split}
\end{equation}
We can solve this equation by introducing an ansatz for $G_{BB(2)}^{\nu \beta}$ as a linear combination of all possible two index tensors ($X^\mu X^\nu$, $\tilde{P}^\mu \tilde{P}^\nu$, $X^\mu \tilde{P}^\nu$, $X^\nu \tilde{P}^\mu$ and  $\eta^{\mu \nu}$). The coefficients are scalar functions of $(X,\tilde{P})$, which include  $X^2$, $X\cdot \tilde{P}$, $\tilde{P}^2$ and their powers. On account of the source terms --  the $X^2$ term from the metric determinant and the leading solution $G_{BB(1)}^{\nu \beta}$ -- the most general ansatz for $G_{BB(2)}^{\nu \beta}$ needed to solve \ref{eomoffshell} takes the form 
\begin{equation}\label{subleadingBB}
\begin{split}
 G_{BB(2)}^{\nu \beta}(\tilde{P},X) &= \frac{e^{i\tilde{P} \cdot X}}{\tilde{P}^2 + M^2} \Bigg( C(\tilde{P},X) \eta^{\nu \beta} + X^{\nu} X^{\beta} + D(\tilde{P},X)\;X^{\nu} \tilde{P}^{\beta} \\&+ E(\tilde{P},X)\;X^{\beta} \tilde{P}^{\nu} + F(\tilde{P},X)\; \tilde{P}^{\nu} \tilde{P}^{\beta}  + \frac{X \cdot X}{2} \;\Big(\eta^{\nu \beta} + \frac{\tilde{P}^{\nu}\tilde{P}^{\beta}}{M^2}\Big)\Bigg)
\end{split}
\end{equation}
with $C(\tilde{P},X),\;D(\tilde{P},X),\; E(\tilde{P},X)$ and $F(\tilde{P},X)$ the coefficients. Using \ref{eomoffshell}, we find the following coefficient solutions

\begin{equation}\label{etaco}
\begin{split}
C(\tilde{P},X) = -\frac{(\tilde{P} \cdot X)^2}{(\tilde{P}^2 + M^2)} &+ 
\frac{i (\tilde{P} \cdot X)\big((2 + d) M^4 + (-3 + d) M^2 \tilde{P}^2 - \tilde{P}^4\big)}{M^2(\tilde{P}^2 + M^2)^2}\\& +\frac{\big(-8 M^2 \tilde{P}^2 + d (M^4 - \tilde{P}^4)\big)}{(\tilde{P}^2 + M^2)^3}
    \end{split}
\end{equation}
\begin{equation}\label{PPco}
\begin{split}
F(\tilde{P},X)= - \frac{(\tilde{P} \cdot X)^2}{M^2(\tilde{P}^2 + M^2)} &+
   \frac{i (\tilde{P} \cdot X)\big((7 + d) M^2 + (3 + d) \tilde{P}^2\big)}{M^2 (\tilde{P}^2 + M^2)^2} \\&+ \frac{(8 + 3 d) M^4 + 4 d M^2 \tilde{P}^2 + d \tilde{P}^4}{M^2 (\tilde{P}^2 + M^2)^3}
    \end{split}
\end{equation}
\begin{equation}\label{sympp}
D(\tilde{P},X)=\frac{(-i + \tilde{P} \cdot X)}{M^2}= E(\tilde{P},X)
\end{equation}

Substituting \ref{etaco}-\ref{sympp} in \ref{subleadingBB}, we get the desired bulk-to-bulk vector propagator solution up to subleading $1/R^{2}$ corrections. In the following section, we discuss how the propagators computed here can be used with Witten diagrams to formulate an AdS S-matrix. We consider the Abelian Higgs model on AdS spacetimes, which provides an example of an exchange process with massive external vector fields.

 \section{AdS S-Matrix}\label{brief}

\begin{figure}[h!]
  \centering
  \begin{tikzpicture}[scale=1.0]
   
    \draw[thin, darkgray!100, line width=0.3mm] (0,0) circle (2.5);
 
    \draw[thin, darkgray!100, line width=0.2mm] (0,0) circle (1);
  
   \draw[thick,dotted] (2,0) -- (1.5,0) arc(360:180:1.5);
    
    \coordinate (1) at (-1.75,1.75);
    \coordinate (2) at (-2.5,0);
    \coordinate (3) at (2.5, 0);
   
    \fill (1) circle (2pt) node[below] {};
    \fill (2) circle (2pt) node[below] {};
    \fill (3) circle (2pt) node[below] {};

      \draw[line width=0.4mm] (1) -- (-0.9,0.4);
      \draw[line width=0.4mm] (2) -- (-1,0);
      \draw[line width=0.4mm] (3) -- (1,0);
 
    \node[left] at (1) {\textcolor{darkgray}{$\tilde{\Pi}(P_1,X_1)$}};
    \node[left] at (2) {\textcolor{darkgray}{$\tilde{\Pi}(P_2,X_2)$}};
    \node[right] at (3) {\textcolor{darkgray}{$\tilde{\Pi}(P_n,X_n)$}};
    \end{tikzpicture}
  \caption{The Witten diagram for a n-point function involving n bulk-to-boundary propagators and a generic bulk interaction (indicated by a circle)}
  \label{fig2}
\end{figure}
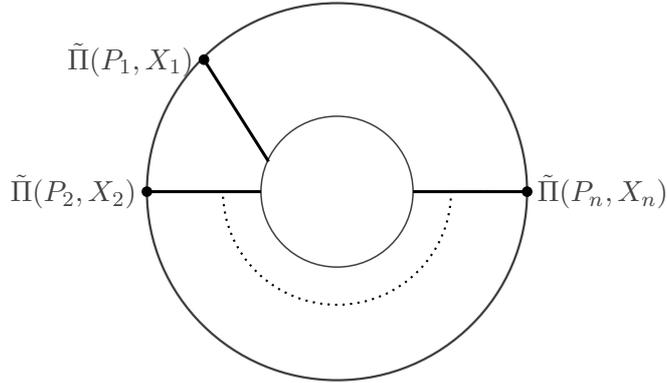

Following the prescription for massive scalar fields in \cite{Gadde:2022ghy}, in this section we will define an ``AdS S-matrix" involving external massive vector fields. To this end, we consider a Witten diagram with $n$ external particles and a generic bulk interaction

\begin{equation}
D(P_1\,,\cdots \,,P_n) = \left(\prod_{i=1}^n \int  d^{d+1} X_i \, \epsilon_{\mu_i} (P_i\,, X_i)\, \Pi^{\mu_i \nu_i} (P_i\,,X_i) \right) B_{\nu_1 \cdots \nu_{n}}(P_1\,, X_1;\cdots\,; P_n\,, X_n)\,,
\label{gen.int}
\end{equation}

with $\epsilon_{\mu_i} (P_i\,, X_i)$ and $\Pi^{\mu_i \nu_i} (P_i\,,X_i)$ being the polarizations and bulk-to-boundary propagators for the external massive vector fields, and $B_{\nu_1 \cdots \nu_{n}}(P_1\,, X_1;\cdots\,; P_n\,, X_n)$ as a generic bulk contribution that will involve interaction vertices and bulk-to-bulk propagators depending on the process under consideration as depicted in Fig. \ref{fig2}.
Having discussed the $1/R^{2}$ corrections to the propagators and the metric previously, we now discuss the derivation for the polarization vectors. This follows from the exact transversality condition satisfied by the massive vector field about the AdS background \ref{lc}

\begin{equation}
\nabla_{\mu} A^{\mu}(X)=\partial_\nu A^\nu(X) - \frac{1}{R^2} X^\nu \eta_{\nu \rho}A^\rho(X)= 0.
\label{lc.r2}
\end{equation}

We can now consider the Fourier transformed massive vector field, that can be expanded to all orders in $1/R^{2}$ 

\begin{equation}
A_{\mu}(X) = \int d^dP \left(\epsilon^{(0)}_{\mu} (P,X) + \frac{1}{R^2} \epsilon^{(1)}_{\mu} (P,X) \right) e^{i P\cdot X}
\label{mv.ft}
\end{equation}

Substituting \ref{mv.ft} in \ref{lc.r2}, we find

\begin{equation}
i P^{\mu}\epsilon^{(0)}_{\mu} + \frac{1}{R^2}\left(i P^{\mu}\epsilon^{(1)}_{\mu} - X^{\mu}\epsilon^{(0)}_{\mu}\right)= 0
\label{pol.m}
\end{equation}
The leading condition $P^{\mu}\epsilon^{(0)}_{\mu}= 0$ is the usual flat spacetime transversality condition, and will depend on the orientation of the massive vector field $\epsilon^{(0)}_{\mu} = \epsilon^{(0)}_{\mu}(\hat{P})$. The $1/R^{2}$ correction in \ref{pol.m} can have a general solution
\begin{equation}
\epsilon^{(1)}_{\mu}(P,X) =  \left(-i\epsilon^{(0)}_{\nu} X^{\nu}\right) \left(-\frac{c_1}{M^2} P_{\mu} + \frac{c_2}{P \cdot X} X_{\mu} \right).
\label{pol1.sol1}
\end{equation} 
Contracting with $P^{\mu}$ and using $P^2 = -M^2$, we note that $c_1 + c_2 := 1$. As we will only consider a solution that involves terms $(P\cdot X)^n$ with $n \ge 1$, we will set $c_2 = 0$ and $c_1 = 1$. Hence the relevant solution up to $1/R^2$ corrections for the polarization vector is 

\begin{equation}
\epsilon_{\mu}(P,X)= \epsilon^{(0)}_{\mu}(\hat{P}) + \frac{1}{R^2}\epsilon^{(1)}_{\mu}(P,X) =  \epsilon^{(0)}_{\mu}(\hat{P}) + i \epsilon^{(0)}_{\nu}(\hat{P}) X^{\nu} \frac{P_{\mu}}{M^2 R^2} 
\label{pol.sol}
\end{equation} 

We now discuss the general approach towards evaluating \ref{gen.int}. The external bulk-to-boundary propagators each come with plane wave contributions $e^{i P_i \cdot X_i}$ and coefficients that depend on $X$. These are broadly of two types, those that depend on the specific combination $P\cdot X$ and its powers, and those that depend on $X^{\mu}$ in any other way. We discuss these cases in turn. For the bulk contribution, we have vertices and bulk-to-bulk propagators with momenta that are integrated over. Following \cite{Gadde:2022ghy}, we inject dual auxiliary momenta at each of the vertices and along each of the bulk-to-bulk propagators.  Formally, this is a consequence of breaking translation invariance in the bulk and will be important in identifying the AdS S-matrix. We first comment on the role of injected momenta and rescaling in performing Witten diagram integrals. With the injected momenta, we resolve terms involving $X_i$ in terms of derivatives with respect to the auxilliary momenta. For instance, for a three point vertex $A(P_1,P_2,P_3,X)$ at bulk position $X$, we express it as
\begin{equation}
A(P_1,P_2,P_3,X) \equiv A(P_1,P_2,P_3,X) e^{i \mathfrak{p} \cdot X}\Big\vert_{\mathfrak{p} = 0} \equiv A\left(P_1,P_2,P_3,-i \partial_{\mathfrak{p}}\right) e^{i \mathfrak{p} X}\Big\vert_{\mathfrak{p} = 0} \,,
\label{pd1}
\end{equation}

with $\mathfrak{p}^{\mu}$ being the injected auxilliary momentum at the vertex which we set to vanish at the end, and $\partial^{\mu}_{\mathfrak{p}} = \frac{\partial}{\partial \mathfrak{p}_{\mu}}$. For simplicity, we denote this as $\partial_{\mathfrak{p}}$. This step of replacing positions by derivatives with respect to injected momenta can be applied along all contributions in a Witten diagram. For coefficients involving the scalar combination $P_i\cdot X$ and its powers, where $P_i$ is an external momentum and $X$ is any bulk point, a useful prescription is to rescale the plane wave exponent by $\sigma_i$ such that

\begin{equation}
C(P_i\cdot X_i) e^{i P_i\cdot X_i} = C(P_i\cdot X_i) e^{i \sigma_i P_i\cdot X_i}\Big \vert_{\sigma_i = 1} 
= C \left(-i \partial_{\sigma_i}\right) e^{i \sigma_i P_i\cdot X_i}\Big \vert_{\sigma_i = 1}\,,
\label{pd2}
\end{equation}

with $\partial_{\sigma_i}= \frac{\partial}{\partial \sigma_i}$.
The bulk-to-bulk propagators in $1/R$ perturbation theory have higher order poles beyond the simple poles encountered in the flat spacetime limit. For these contributions, we rescale the exchanged particle mass in the propagator $m\rightarrow \alpha m$ and consider derivatives with respect to it. We can thus consider
\begin{equation}
\frac{1}{(P^2 + m^2)^{n+1}} = \frac{(-1)^n}{n! ~m^{2n}} \partial_{\alpha}^n  \frac{1}{P^2 + \alpha m^2} \Big \vert_{\alpha = 1}\,,
\label{pd3}
\end{equation}

where $\partial_{\alpha}^n =\frac{\partial^n}{\partial \alpha^n}$. An explicit consideration of this is provided through our example in the following subsection, and further details behind this construction are discussed in \cite{Gadde:2022ghy}.

Lastly, in considering the bulk integral over positions, there can arise an ambiguity in the final expression for the amplitude in terms of momenta. To see this, we note that

\begin{equation}
    \int d^{d+1}X \; (P \cdot X) \; e^{i P \cdot X} = -i \int d^{d+1}X \; X^\mu \pdv{}{X^\mu}\; e^{i P \cdot X} = i(d+1) \int d^{d+1}X \; e^{i P \cdot X}\,,
    \label{identity}
\end{equation}

with appropriate boundary conditions on $X$ to go to the last equality in \ref{identity}. This ambiguity can be resolved by expressing one of the external momenta, conventionally taken to be $P_n$, as the difference of all other momenta from the total sum over all momenta, i.e. $P_n = \sum_{i=1}^n P_i - \sum_{j=1}^{n-1}P_j$. In this way, the resulting expressions are expressed in terms of $\{P_1\,,\cdots P_{n-1}\}$. In practice, this amounts to expressing the final result in terms of the relevant independent Mandelstam variables.

With the above discussion, we now define the AdS S-matrix from \ref{gen.int}. The injected momenta and rescalings account for $1/R$ bulk corrections that in general break the translation invariance present in the strict $R \to \infty$ limit. The parametrized derivatives in \ref{pd1} - \ref{pd3} act on the complete Witten diagram, and we get two sets of contributions. One of these, which we denote as $A_{(0)}(P_i,\epsilon_i)$, comes from the derivatives acting on all of integrand apart from the plane wave states. The integral is thus proportional to a momentum conserving delta function (from the plane wave pieces) and involve $1/R$ corrections (from the propagators and polarizations). This contribution from the Witten diagram is called the AdS S-matrix. It respects momentum conservation and leads to the flat space S-matrix when $R\rightarrow \infty$ without any ambiguity. 

The derivatives \ref{pd1} - \ref{pd3} also act on the plane wave states and these give derivatives of the momentum conserving delta function. As a result, and on noting the prescription for fixing $P_n$, the evaluation of \ref{gen.int} gives

\begin{equation}
D(P_1\,,\cdots \,,P_n) 
= \Big[A_{(0)}(P_i,\epsilon_i) + [A_{(1)}(P_i,\epsilon_i)]^{\mu_1} \Big(-i \pdv{}{P_n^{\mu_1}}\Big) + \dots \Big] \delta\big(\sum_i P_i\big)
\label{defSmatrix}
\end{equation}

Hence a boundary correlation function has a Witten diagram representation that evaluates into a contribution proportional to a momentum conserving delta function that is called the AdS S-matrix, and a set of terms proportional to derivatives of the momentum delta function that represent the breaking of translation invariance beyond the flat spacetime limit. The AdS S-matrix has two salient features. The first is that they capture relevant $1/R$ corrections of a flat spacetime scattering amplitude in the bulk, in the sense of being proportional to a momentum conserving delta function. This makes it a useful description for local bulk physics as perturbations about flat spacetime. However, the AdS S-matrix can also recover the complete boundary correlation function, i.e. the pieces involving derivatives of the momentum conserving delta function. This follows from substituting it in a conformal Ward identity and using the fact that $D(P_1\,,\cdots \,,P_n)$ is invariant under the action of $M_{AB}$ \cite{Gadde:2022ghy}. 
\begin{equation}
   M_{AB} = \Big(\mathcal{P}_A\pdv{}{\mathcal{P}^B}-  \mathcal{P}_B \pdv{}{\mathcal{P}^A}\Big) \,,\qquad A,B = 0 \,,\cdots d+1 .
\end{equation}
More specifically about the flat patch, we get a non-trivial contribution from the action of $M_{\mu\,, d+1} =  R P_{\mu}$ (where $\mu = 0\,,\cdots d$) on $A_{(0)}(P_i,\epsilon_i)$, from the Ward identity, and it recovers terms proportional to derivatives of the momentum conserving delta function to each order in $1/R^2$ \cite{Gadde:2022ghy}. In this way the AdS S-matrix, along with the conformal Ward identity for $M_{\mu, d+1}$, capture local and global information of the correlation function with external legs on the boundary of the AdS$_{d+1}$ background. In the next  subsection, we will use the above prescription to compute a four-point amplitude in the Abelian Higgs Model.

\subsection{Abelian Higgs Model and exchange interaction}
In this subsection we consider an Abelian Higgs model on $AdS_{d+1}$. This  provides us a concrete example of an interacting Proca theory after the spontaneous breaking of the local $U(1)$ gauge symmetry by a complex scalar field, and we will subsequently consider a four point tree level exchange diagram to compute an AdS S-matrix.  The Abelian Higgs model is important in several areas of theoretical physics in a much broader context. In the following, we will focus on the couplings relevant for the four vector field process using Witten diagrams, and subsequently, its resulting contribution to the AdS S-matrix.

The action for the Abelian Higgs model coupled to a gauge field in curved spacetime is given by:

\begin{equation}\label{action}
S_{\rm matter} = \int d^4x \sqrt{-g} \left(-\frac{1}{4} F^{\mu\nu} F_{\mu\nu} - (D^\mu \phi)^{*} (D_\mu \phi) - V(\phi^*\phi) \right)
\end{equation}
where \( F^{\mu\nu} = \nabla^{\mu} A^{\nu} - \nabla^{\nu} A^{\mu} \) is the electromagnetic field strength tensor, \( D_{\mu} = \partial_{\mu} - i g' A_{\mu} \) is the covariant derivative, \(\phi\) is a complex scalar field, \( A_{\mu} \) is the vector field, and the potential is given by,

\begin{equation}\label{Potential}
 V(\phi,\phi^*) = \mu^2 |\phi|^2 + {\lambda} |\phi|^4  
\end{equation}
One can note here that in the above expression \ref{action}, there is no mass term for the vector field. For $\mu^2<0$ the potential given in \ref{Potential} has two minima at \Big($|\phi| = \pm \sqrt{\frac{- \mu^2}{2\lambda}}$~\Big). Hence we have two degenerate vacua which spontaneously breaks the local $U(1)$ gauge symmetry. Since the complex scalar field can be written as $\phi=|\phi| e^{i\alpha}$, we choose the VEV for the $\phi$ field to be  $\langle \phi(x)\rangle = \phi_0 = \sqrt{\frac{- \mu^2}{2\lambda}}$. We can thereafter rewrite our theory given in \ref{action} by defining $\phi(x) = \rho(x) e^{i \alpha(x)}$ where $\rho(x)=\frac{1}{\sqrt{2}}(\phi_{0}+\Phi(x))$ and with $\Phi(x)$ the Higgs field. A similar matter action was considered in \cite{Dehghani:2001nz,Gubser:2008px} in particular on the AdS$_4$ black hole background.

The action in terms of the field $\Phi(x)$ can be written as,
\begin{equation}\label{fullaction}
\begin{split}
     S[\Phi] = -\int d^4x \sqrt{-g} \Big(\frac{1}{2}\partial^{\mu}\Phi\partial_{\mu}\Phi & + \frac{1}{2}\lambda \phi_0^2 \Phi^2 +\frac{1}{4} F^{\mu\nu} F_{\mu\nu} + \frac{1}{2} g'^2 \phi_0^2 A_{\mu} A^{\mu} + g'^2 \phi_0 \Phi  A_{\mu}A^{\mu}  \\
     &  +\cdots \Big)\,,
\end{split} 
\end{equation}
where we have only considered terms in $\Phi$ and $A_{\mu}$ relevant for our analysis to follow, while $\cdots$ in \ref{fullaction} refer to contributions from the phase $\alpha(x)$, the scalar current coupled to the vector field, as well as the scalar cubic, quartic and constant terms. In the above expression, we see that the vector field acquires a mass once our gauge symmetry is spontaneously broken. From the mass term $ {1\over 2}g'^2 \phi_0^2 A_{\mu} A^{\mu}$ corresponding to the vector boson we can write its mass as,
\begin{equation}
    M = g'\phi_0,
\end{equation}
The mass of the Higgs field $\Phi$ is given by,
\begin{equation}
    m_{\Phi}= \sqrt{\lambda}\phi_0.
\end{equation}

In the following, we illustrate the AdS S-matrix computation for a process resulting from \ref{fullaction} involving four external Proca fields mediated by a massive scalar at tree-level. Since the action in \ref{fullaction} does not involve any quartic or cubic term in $A_{\mu}$, we do not have any contact or exchange diagrams involving  only the vector fields. 

\begin{figure}[h!]
  \centering
  \begin{tikzpicture}[scale=1.0]
    \draw[thin, darkgray!100, line width=0.3mm] (0,0) circle (2.5);

    \coordinate (1) at (-1.75,1.75);
    \coordinate (2) at (-1.75,-1.75);
    \coordinate (3) at (1.75,1.75);
    \coordinate (4) at (1.75,-1.75);
    \fill (1) circle (2pt) node[below] {};
    \fill (2) circle (2pt) node[below] {};
    \fill (3) circle (2pt) node[below] {};
    \fill (4) circle (2pt) node[below] {};

    \draw[decorate, decoration=snake, Darkblue!100!black, line width=0.4mm] (1) -- (-0.8,0);
      \draw[decorate, decoration=snake, Darkblue!100!black, line width=0.4mm] (2) -- (-0.8,0);
     \draw[thick,dashed , Darkblue!100!black, line width=0.4mm] (-0.8,0) -- (0.8,0);
    \draw[decorate, decoration=snake, Darkblue!100!black, line width=0.4mm] (0.8,0)-- (3);
      \draw[decorate, decoration=snake, Darkblue!100!black, line width=0.4mm] (0.8,0)-- (4);
     \node[above] at (0,0) {$G_{BB}$};

 \fill (-0.8,0) circle (0.08);
 \fill (0.8,0) circle (0.08);
   
    \node[left] at (1) {\textcolor{darkgray}{$\mathcal{O}_1(P_1)$}};
    \node[left] at (2) {\textcolor{darkgray}{$\mathcal{O}_2(P_2)$}};
    \node[right] at (3) {\textcolor{darkgray}{$\mathcal{O}_3(P_3)$}};
    \node[right] at (4) {\textcolor{darkgray}{$\mathcal{O}_4(P_4)$}};
\node[above] at (-1.4,1.4) {$\alpha$};
\node[above] at (-1.4,-2) {$\beta$};
\node[above] at (1.4,1.4) {$\gamma$};
\node[above] at (1.4,-2) {$\delta$};

    \node[below] at (-1.3,0.2) {$\nu$};
      \node[below] at (-0.7,-0.1) {$X$};
    \node[below] at (1.3,0.2) {$\rho$};
    \node[below] at (0.65,-0.1) {$Y$};
  \end{tikzpicture}
  \caption{Witten diagram representing a four-point exchange diagram mediated by a Higgs field $\Phi$. The external legs correspond to vector fields $A^{\alpha}$, $A^{\beta}$, $A^{\gamma}$ and $A^{\delta}$ with momenta $P_1$, $P_2$, $P_3$ and $P_4$ respectively. $\mathcal{O}_i(P_i)$s denote their dual boundary CFT operators.}
  \label{fig:my_label2}
\end{figure}
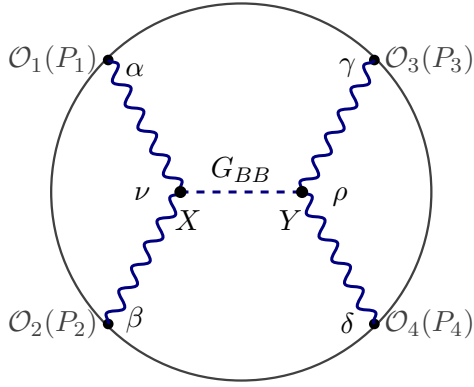

We now consider a tree-level exchange Witten diagram involving four external massive vector fields $A^{\alpha}$, $A^{\beta}$, $A^{\gamma}$ and $A^{\delta}$ with momenta $P_1$, $P_2$, $P_3$ and $P_4$ respectively with mass $M$, exchanging a massive scalar $\Phi$ with mass $m_{\Phi}$. The $3$-point interaction vertex is determined from the term $\Phi  A_{\mu} A^{\mu}$ in the action \ref{fullaction}, and where for notational simplicity we set $g'^2 \phi_0 = 1$. We denote the corresponding bulk vertices by $X$ and $Y$ in position space. The Witten diagram for this process is given in Fig. \ref{fig:my_label2}.

\par

It has the integral representation
\begin{equation}\label{abhiggs}
\begin{split}
D(\{P_i, M, m_{\Phi}\})&=\int dX \sqrt{-g(X)}\int dY\sqrt{-g(Y)} ~ \epsilon_{\alpha}(P_1,X)\epsilon_{\beta}(P_2,X)\epsilon_{\gamma}(P_3,Y)\epsilon_{\delta}(P_4,Y)\\
& g_{\mu\nu}(X)\tilde{\Pi}^{\mu\alpha}(P_1,X)\tilde{\Pi}^{\nu\beta}(P_2,X) ~ g_{\rho\sigma}(Y)\tilde{\Pi}^{\rho\gamma}(P_3,Y)\tilde{\Pi}^{\sigma\delta}(P_4,Y)  G_{BB}(X,Y) \,,
\end{split}
\end{equation}
where $\epsilon_{\alpha}(P_1\,,X), \cdots ,\epsilon_{\delta}(P_4\,,Y)$ are the polarization vectors introduced in \ref{pol.sol} for the four external vector fields with bulk-to-boundary propagators $\tilde{\Pi}^{\alpha\mu}_1(P_1,X), \cdots ,\tilde{\Pi}^{\delta\sigma}_4(P_4,Y)$ given in \ref{fullbdyP}, $g_{\mu \nu}$ is the metric from \ref{met}, $\sqrt{-g}$ is its determinant given by

 \begin{equation}
 \sqrt{-g(X)} = 1-\frac{X^2}{2 R^2}+ \mathcal{O}(R^4)\,,
 \end{equation}
 
 and $G_{BB}(X,Y)$ is the bulk-to-bulk scalar propagator connecting the bulk interaction vertices $X$ and $Y$.  The bulk-to-bulk propagator for a scalar field with a mass $m_{\Phi}$, and with one position Fourier transformed was derived in \cite{Gadde:2022ghy} and has the following expression

 \begin{equation} \label{massscalar}
    \begin{aligned}
G_{\rm BB}(P\,; Y) & =\frac{e^{i{P}\cdot Y}}{{P}^2+m_{\Phi}^2}\Bigg[ 1 +\frac{1}{R^2}\left(\frac{Y^2}{2}+\frac{-({P}\cdot Y)^2+i(d-1) {P}\cdot Y}{\left({P}^2+m_{\Phi}^2\right)}\right. \\
& \left.+\frac{4 i m_{\Phi}^2({P} \cdot Y)-(d-1) {P}^2}{\left({P}^2+m_{\Phi}^2\right)^2}+\frac{(d-7) m_{\Phi}^2{P}^2+(d+1) m_{\Phi}^4}{\left({P}^2+m_{\Phi}^2\right)^3}\right) \Bigg] + \mathcal{O}(1/R^4) \\
\end{aligned}
\end{equation}

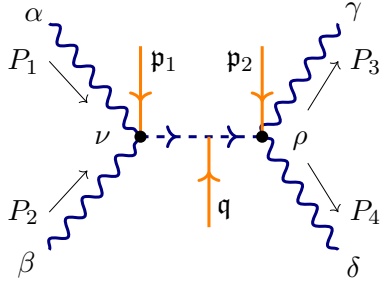
\begin{figure}[hbt]
  \centering
  \begin{tikzpicture}
   \draw[decorate, decoration=snake, Darkblue!100!black, line width=0.4mm] (1) (-2,1.5) -- (-0.8,0);
      \draw[decorate, decoration=snake, Darkblue!100!black, line width=0.4mm] (2) (-2,-1.5) -- (-0.8,0);
      
     \draw[thick, dashed, Darkblue!100!black, line width=0.4mm, postaction={decorate, decoration={markings, mark=at position 0.3 with {\arrow{>}}}}, decoration={markings, mark=at position 0.8 with {\arrow{>}}}] (3)(-0.8,0) -- (0.8,0);
  
 \draw[thick, orange!100!black, line width=0.4mm, postaction={decorate, decoration={markings, mark=at position 0.5 with {\arrow{<}}}}] (4) (0.1,0) -- (0.1,-1.2);
 
\draw[decorate, decoration=snake, Darkblue!100!black, line width=0.4mm](0.8,0)-- (1.8,1.5);
      \draw[decorate, decoration=snake, Darkblue!100!black, line width=0.4mm] (0.8,0)-- (1.8,-1.5);
   \node[right] at (-0.8,1) {$\mathfrak{p}_1$};
      \node[below] at (0.3,-0.7) {$\mathfrak{q}$};
      \node[right] at (0.2,1) {$\mathfrak{p}_2$};
     \draw[thick, orange!100!black, line width=0.4mm , postaction={decorate, decoration={markings, mark=at position 0.5 with {\arrow{<}}}}] (-0.8,0) -- (-0.8,1.2);
        \draw[thick, orange!100!black, line width=0.4mm , postaction={decorate, decoration={markings, mark=at position 0.5 with {\arrow{<}}}}] (0.8,0) -- (0.8,1.2);
 \fill (-0.8,0) circle (0.08);
  \fill (0.8,0) circle (0.08);

\draw [->, thick,black!100!black, line width=0.1mm,] (-2,1) -- (-1.5,0.45);
\draw [->, thick,black!100!black, line width=0.1mm,] (-2,-1) -- (-1.5,-0.45);
\draw [<-, thick,black!100!black, line width=0.1mm,] (1.8,1) --(1.4,0.35);
\draw [<-, thick,black!100!black, line width=0.1mm,] (1.8,-1) -- (1.4,-0.35);

 \node[left] at(-2,1) {$P_1$};
     \node[left] at (-2,-1) {$P_2$};
      \node[left] at(2.5,1) {$P_3$};
            \node[left] at(2.5,-1) {$P_4$};
\node[above] at (-2.2,1.4) {$\alpha$};
\node[above] at (-2.3,-2) {$\beta$};
\node[above] at (2,1.4) {$\gamma$};
\node[above] at (2,-2) {$\delta$};
 \node[below] at (1.3,0.2) {$\rho$};
    \node[below] at (-1.3,0.2) {$\nu$};
  \end{tikzpicture}
  \caption{Higgs exchange diagram with momentum injections. Yellow lines represent the injected momenta $\mathfrak{p}_1$ and $\mathfrak{p}_2$ at the vertices and $\mathfrak{q}$ injection at the propagator.}
  \label{fig:momentainjection}
\end{figure}

We can now derive the AdS S-matrix from the relevant Witten diagram as shown in Fig.\ref{fig:momentainjection}. We inject momenta $\mathfrak{p_1}$ and $\mathfrak{p_2}$ at the vertices at $X$ and $Y$ respectively, along with $\mathfrak{q}$ in the bulk-to-bulk propagator. 

We apply the prescription of the previous section on \ref{abhiggs} and address this for individual terms appearing in the integrand, beginning with the bulk-to-bulk propagator $G_{BB}(X,Y)$. We parametrize the mass in the propagator \ref{massscalar} as $\alpha m_{\Phi}^2$, with which terms with $(P^2 + m_{\Phi}^2)^{-n}$ for $n \ge 2$ can be described in terms of derivatives with respect to $\alpha$ of $(P^2 + \alpha m_{\Phi}^2)^{-1}$ and taking $\alpha = 1$ at the end. More specifically, we can express \ref{massscalar} as

\begin{align}
G_{\rm BB}(P\,; Y)  = g_{\rm BB}(Y^2, P\cdot Y, \partial_{\alpha}) \frac{e^{i{P}\cdot Y}}{{P}^2+ \alpha m_{\Phi}^2} \Bigg\vert_{\alpha = 1}
\end{align}
where
\begin{align}
g_{\rm BB}(Y^2, P\cdot Y, \partial_{\alpha}) &= 1 + \frac{1}{R^2} \left( \frac{Y^2}{2} - \frac{- (P\cdot Y)^2 + i (d-1) P\cdot Y}{m_{\Phi}^2} \partial_{\alpha} \right. \notag\\
&.\left. + \frac{4 m_{\Phi}^2 (P\cdot Y) - (d-1) P^2}{2 m_{\Phi}^4} \partial_{\alpha}^2 - \frac{(d-7) m_{\Phi}^2 P^2 + (d+1) m_{\Phi}^4}{6 m_{\Phi}^6} \partial_{\alpha}^3 \right) \,.
\label{gbb.def}
\end{align}

The notation $g_{\rm BB}(Y^2, P\cdot Y, \partial_{\alpha})$ indicates the dependence on $Y^2\,, P\cdot Y$ and $\partial_{\alpha}$, which will be relevant in deriving the AdS S-matrix from Witten diagrams.
 The injected momentum $\mathfrak{q}$ along the bulk-to-bulk propagator is realized through the term $e^{i \mathfrak{q}\cdot Y}$, which allows us to replace all dependence on $Y$ with $-i\partial_{\mathfrak{q}}$. Hence \ref{gbb.def} has the equivalent description

\begin{equation}
G_{\rm BB}(P\,; Y)  = g_{\rm BB}((-i\partial_{\mathfrak{q}})^2, -i P\cdot \partial_{\mathfrak{q}}, \partial_{\alpha}) \frac{e^{i{P}\cdot Y} e^{i \mathfrak{q}\cdot Y}}{{P}^2+ \alpha m_{\Phi}^2} \Bigg\vert_{\alpha = 1, \mathfrak{q} = 0}
\label{gbb.aq}
\end{equation}

Lastly, we have $G_{BB}(X,Y)$ from the inverse Fourier transform of \ref{gbb.aq} giving

\begin{equation}
G_{\rm BB}(X\,; Y)  = \int dP \, g_{\rm BB}((-i\partial_{\mathfrak{q}})^2, -i P\cdot \partial_{\mathfrak{q}}, \partial_{\alpha}) \frac{e^{i{P}\cdot (Y- X)} e^{i \mathfrak{q}\cdot Y}}{{P}^2+ \alpha m_{\Phi}^2} \Bigg\vert_{\alpha = 1, \mathfrak{q} = 0}
\label{gbb.xy}
\end{equation}

Apart from the bulk-to-bulk propagator, we have the three point vertex, metric determinant and polarization contributions at $X$ and $Y$. In the following, we discuss the contribution at $X$ in detail and state the result at $Y$. From \ref{abhiggs}, we have the following terms at bulk point $X$, 

\begin{align}
\sqrt{-g (X)} \epsilon_{\alpha}(P_1\,,X) \epsilon_{\beta}(P_2\,,X) V_{12}^{\alpha \beta}(P_1\cdot X\,, P_2\cdot X\,, X) e^{i P_1\cdot X}e^{i P_2\cdot X}
\label{vp1}
\end{align}

where we introduced the following notation to describe a three-vertex 
\begin{align}
V_{ij}^{\alpha \beta}(P_i\cdot Z\,, P_j\cdot Z\,, Z) e^{i P_i\cdot Z}e^{i P_j\cdot Z}
= g_{\mu\nu}(Z)\tilde{\Pi}^{\mu\alpha}(P_i,Z)\tilde{\Pi}^{\nu\beta}(P_j,Z) 
\end{align}

which indicates the relevant dependence on $P_i\cdot Z\,, P_j\cdot Z$ and $Z$ for a vertex at $Z$ and $i,j$ denote the external legs at the vertex. More specifically, using \ref{met} and \ref{fullbdyP}, we find the contribution at $X$ apart from the polarizations to be

\begin{align}
&\sqrt{-g (X)} V_{12}^{\alpha \beta}(P_1\cdot X\,, P_2\cdot X\,, X) \notag\\&=  G^{\mu \alpha}(P_1) G^{\nu \beta}(P_2) \left[\left(1 - \frac{X^2}{2 R^2}\right)\eta_{\mu \nu} - \frac{X_{\mu} X_{\nu}}{R^2}\right] \notag\\
& \quad + \eta_{\mu \nu}\frac{G^{\nu \beta}(P_2)}{M^2 R^2} \left(i(d+1) X^{\alpha}{P_1}^{\mu} -i X^{\mu} {P_1}^{\alpha} - \frac{d}{M^2} {P_1}^{\alpha} {P_1}^{\mu}\right) \notag\\
& \qquad + \eta_{\mu \nu}\frac{G^{\mu \alpha}(P_1)}{M^2 R^2} \left(i(d+1) X^{\beta}{P_2}^{\nu} -i X^{\nu} {P_2}^{\beta} - \frac{d}{M^2} {P_2}^{\beta} {P_2}^{\nu}\right)    \notag\\
& \qquad \qquad  + \eta_{\mu \nu}\frac{G^{\mu \alpha}(P_1) G^{\nu \beta}(P_2)}{M^2 R^2}\left(\frac{i(d+2)}{4} \left(({P_1}\cdot X)  + ({P_2}\cdot X)\right) +\frac{d}{4}\left(({P_1}\cdot X)^2 + ({P_2}\cdot X)^2\right) \right.  \notag\\
& \left. \qquad \qquad \qquad \qquad \qquad \qquad \qquad  \quad + \frac{i}{6}\left(({P_1}\cdot X)^3 + ({P_2}\cdot X)^3\right)\right)
\label{v12.exp}
\end{align}
where $G^{\mu \alpha}(P)=\eta^{\mu\alpha}-\frac{P^{\mu}P^{\alpha}}{|P|^2}$ and we have kept terms up to $1/R^2$ order. 

The $1/R^{2}$ correction of the polarization vector given in \ref{pol.sol} does not contribute in \ref{abhiggs}. To see this, we use \ref{pol.sol} and \ref{v12.exp} in \ref{vp1} to find

\begin{align}
&\epsilon_{\alpha}(P_1\,,X) \Big \vert_{1/R^{2}} \sqrt{-g (X)}   V_{12}^{\alpha \beta}(P_1\cdot X\,, P_2\cdot X\,, X) e^{i (P_1+P_2)\cdot X}\Big\vert_{R \to \infty} \notag\\
&= i \epsilon^{(0)}_{1,\gamma} X^{\gamma} \frac{(P_1){}_{\alpha}}{M^2 R^2}  \left( G^{\mu \alpha} (P_1) e^{i P_1.X} G^{\nu \beta} (P_2) e^{i P_2.X} \right) = 0\,,
\end{align}

where we used the property $P_{1 \alpha} G^{\mu \alpha}(P_1) = 0$. We also introduced the shorthand notation $\epsilon_{\alpha}^{(0)}(\hat{P}_i) = \epsilon_{i, \alpha}^{(0)}$ to simplify the notation for polarization vectors for the external particles. Hence we replace all polarizations appearing in \ref{abhiggs} by their leading $R \to \infty$ contribution and accordingly write \ref{vp1} as
\begin{align}
\sqrt{-g (X)} \epsilon^{(0)}_{1,\alpha}\epsilon^{(0)}_{2,\beta} V_{12}^{\alpha \beta}(P_1\cdot X\,, P_2\cdot X\,, X) e^{i P_1.X}e^{i P_2.X}
\label{vp2}
\end{align}

Lastly, we inject the momentum $\mathfrak{p}_1^{\mu}$ by introducing $e^{i \mathfrak{p}_1\cdot X}$ at the vertex, with which we replace the dependence on $X$ (apart from $P\cdot X$ terms) with $-i\partial_{\mathfrak{p}_1}$. We also rewrite $e^{i P_1\cdot X}$ and $e^{i P_2\cdot X}$ as $e^{i \sigma_1 P_1\cdot X}\vert_{\sigma_1 = 1}$ and $e^{i \sigma_2 P_2\cdot X}\vert_{\sigma_2 = 1}$ as in \ref{pd2}. This allows us to replace all $X^{\mu}\,, P_1\cdot X$ and $P_2\cdot X$ terms in \ref{vp2} by $-i \partial_{\mathfrak{p}_1}\,, -i \partial_{\sigma_1}$ and $-i\partial_{\sigma_2}$ respectively. Hence

\begin{align}
&\sqrt{-g (X)} \epsilon_{\alpha}(P_1\,,X) \epsilon_{\beta}(P_2\,,X) V_{12}^{\alpha \beta}(P_1\cdot X\,, P_2\cdot X\,, X) e^{i P_1\cdot X}e^{i P_2\cdot X} \notag\\
& \quad =\sqrt{-g (X)} \epsilon^{(0)}_{1,\alpha} \epsilon^{(0)}_{2,\beta} V_{12}^{\alpha \beta}(P_1\cdot X\,, P_2\cdot X\,, X) e^{i \sigma_1 P_1\cdot X}e^{i \sigma_2 P_2\cdot X} e^{i \mathfrak{p}_1\cdot X} \Big\vert_{\sigma_1 = 1 = \sigma_2\,, \mathfrak{p}_1 = 0} \notag\\ 
&\qquad = \sqrt{-g (-i\partial_{\mathfrak{p}_1})} \epsilon^{(0)}_{1,\alpha} \epsilon^{(0)}_{2,\beta} V_{12}^{\alpha \beta}(-i\partial_{\sigma_1}\,, -i\partial_{\sigma_2}\,, -i\partial_{\mathfrak{p}_1}) e^{i X.(\mathfrak{p}_1 + \sigma_1 P_1 + \sigma_2 P_2)} \Big \vert_{\sigma_1 = 1 = \sigma_2\,, \mathfrak{p}_1 = 0} 
\label{vpx}
\end{align}

An identical analysis for the contribution from vertex $Y$ in \ref{abhiggs} can be carried out and we have

\begin{align}
&\sqrt{-g (Y)} \epsilon_{\gamma}(P_3\,,Y) \epsilon_{\delta}(P_4\,,Y) V_{34}^{\gamma \delta}(P_3.Y\,, P_4.Y\,, Y) e^{i P_3.Y}e^{i P_4.Y} \notag\\
&\qquad = \sqrt{-g (-i\partial_{\mathfrak{p}_2})} \epsilon^{(0)}_{3,\gamma} \epsilon^{(0)}_{4,\delta} V_{34}^{\gamma \delta}(-i\partial_{\sigma_3}\,, -i\partial_{\sigma_4}\,, -i\partial_{\mathfrak{p}_2}) e^{i Y.(\mathfrak{p}_2 + \sigma_3 P_3 + \sigma_4 P_4)} \Big \vert_{\sigma_3 = 1 = \sigma_4\,, \mathfrak{p}_2 = 0}
\label{vpy}
\end{align}

Since $\epsilon^{(0)}_{1,\alpha} \epsilon^{(0)}_{2,\beta} \epsilon^{(0)}_{3,\gamma} \epsilon^{(0)}_{4,\delta}$ neither contain any derivatives nor any dependence on the integration variables introduced in \ref{abhiggs}, we will simply consider the computation of $D^{\alpha\beta\gamma\delta}(\{P_i, M, m_{\Phi}\})$ where
\begin{align}\label{amp}
    D(\{P_i, M, m_{\Phi}\})= \epsilon^{(0)}_{1,\alpha} \epsilon^{(0)}_{2,\beta} \epsilon^{(0)}_{3,\gamma} \epsilon^{(0)}_{4,\delta}D^{\alpha\beta\gamma\delta}(\{P_i, M, m_{\Phi}\})\,,
\end{align}

with
\begin{equation}\label{abhiggs1}
\begin{split}
&D^{\alpha\beta\gamma\delta}(\{P_i, M, m_{\Phi}\})=\int  dX dY d{{P}}\sqrt{-g(-i\partial_{\mathfrak{p}_1})}\\
& \Big\{\sqrt{-g(-i\partial_{\mathfrak{p}_2})} V_{12}^{\alpha\beta}(-i\partial_{\sigma_1},-i\partial_{\sigma_2},-i\partial_{\mathfrak{p}_1})V_{34}^{\gamma\delta}(-i\partial_{\sigma_3},-i\partial_{\sigma_4},-i\partial_{\mathfrak{p}_2}){g}_{BB}(-\partial_\mathfrak{q}^2,-i{{P}}\cdot\partial_\mathfrak{q},\partial_{\alpha})\\& \frac{1}{{{P}}^2+\alpha m_{\Phi}^2} e^{iX\cdot (-{{P}}+\sigma_1 P_1+\sigma_2 P_2+\mathfrak{p}_1)}e^{iY\cdot ({{P}}+\mathfrak{q}+\sigma_3 P_3+\sigma_4 P_4+\mathfrak{p}_2)}\Big\} \Big|_{\mathfrak{p}_1=\mathfrak{p}_2=\mathfrak{q}=0; \sigma_i= 1 = \alpha}\,,
\end{split}
\end{equation}
where $i=1,\dots, 4$. We will compute \ref{abhiggs1} and contract with the polarizations at the end to get \ref{amp}. This will provide us the AdS S-matrix using \ref{defSmatrix}. At this point, we make some important observations about \ref{abhiggs1}. All derivative operators appearing at the vertices and the exchange propagator appear with a $R^{-2}$ factor. In general, we would need to respect the ordering of the individual operators before moving all the plane wave contributions to either the left or the right. However, since we are interested in terms up to $R^{-2}$, this ordering issue does not appear in our considerations, leading to the expression as given in  \ref{abhiggs1}.

We can now perform the integrals. The $X$ integral gives $\delta(-P+\sigma_1 P_1+\sigma_2 P_2+\mathfrak{p}_1)$. Subsequently, the integral over $P$ replaces $P^{\alpha}$ with $\Pi^{\alpha}=\sigma_1 P_1^{\alpha}+\sigma_2 P_2^{\alpha}+\mathfrak{p}_1^{\alpha}$. Lastly, performing the integral over $Y$ yields the result

\begin{equation}\label{abhiggs.1}
\begin{split}
&D^{\alpha\beta\gamma\delta}(\{P_i, M, m_{\Phi}\})=\Big\{ \sqrt{-g(-i\partial_{\mathfrak{p}_1})}\sqrt{-g(-i\partial_{\mathfrak{p}_2})}\\
& V_{12}^{\alpha\beta}(-i\partial_{\sigma_1},-i\partial_{\sigma_2},-i\partial_{\mathfrak{p}_1})V_{34}^{\gamma\delta}(-i\partial_{\sigma_3},-i\partial_{\sigma_4},-i\partial_{\mathfrak{p}_2}) {g}_{BB}(-\partial_\mathfrak{q}^2,-i{{P}}\cdot\partial_\mathfrak{q},\partial_{\alpha})\\& \qquad \frac{\delta(\sigma_1 P_1+\sigma_2 P_2+\sigma_3 P_3+\sigma_4 P_4+\mathfrak{p}_1+\mathfrak{p}_2+\mathfrak{q})}{{\Pi}^2+\alpha m_{\Phi}^2} \Big\}\Big|_{\mathfrak{p}_1=\mathfrak{p}_2=\mathfrak{q}=0; \sigma_i= 1 = \alpha}\,.
\end{split}
\end{equation}

The operator in the first line of \ref{abhiggs.1} acts on the second line. The AdS S-matrix contribution is simply that which acts on all terms other than the delta function, and corresponds to the first term on the right hand side of \ref{defSmatrix}. We also recall from the discussion after \ref{defSmatrix}, which follows \cite{Gadde:2022ghy}, that terms involving derivatives of the momentum conserving delta function can be recovered from the Ward identity for $M_{\mu, {d+1}}$. As a result, we find

\begin{equation}\label{abhiggs2}
    \begin{split}
         &D^{\alpha\beta\gamma\delta}(\{P_i, M, m_{\Phi}\})  = A_{(0)}^{\alpha\beta\gamma\delta}(\{P_i, M, m_{\Phi}\})+ \text{terms with derivatives of delta function} \notag\\
        &A_{(0)}^{\alpha\beta\gamma\delta}(\{P_i, M, m_{\Phi}\})\notag\\ & = \delta\left(\sum_{i=1}^4 P_i \right) \Big\{\sqrt{-g(-i\partial_{\mathfrak{p}_1})} V_{12}^{\alpha\beta}(-i\partial_{\sigma_1},-i\partial_{\sigma_2},-i\partial_{\mathfrak{p}_1})V_{34}^{\gamma\delta}(0,0,0){g}_{BB}(0,0,\partial_{\alpha}) \frac{1}{\Pi^2+\alpha m_{\Phi}^2}\Big\}\Bigg|_{L} 
    \end{split}
\end{equation}

 On factoring out the delta function and taking the limits on the parametrizations for the plane waves and bulk propagators, as well as the injected momenta, we recover the usual momentum conserving delta function for the process. Hence, the $A^{\alpha\beta\gamma\delta}_{(0)}(\{P_i, M, m_{\Phi}\})$ in \ref{abhiggs} is simply the AdS S-matrix term $A_{(0)}(P_i\,,\epsilon_i)$ in \ref{defSmatrix} up to contractions with the polarization vectors. Additionally, by factoring out the momentum conserving delta function we simplify the dependence on all derivative terms appearing inside the parenthesis of \ref{abhiggs2} as they only act on $\frac{1}{\Pi^2+\alpha m_{\Phi}^2}$. As a result, we have a dependence on $\{\sigma_1\,, \sigma_2\,, \alpha\,,\mathfrak{p}_1\}$, and we have denoted the limit for these terms by $\vert_L = \vert_{\sigma_1 = 1 = \sigma_2 \,, \alpha = 1\,, \mathfrak{p}_1 = 0}$, which will be used at the end of the calculation. 

It is now a straightforward exercise to compute $A_{(0)}^{\alpha\beta\gamma\delta}(\{P_i, M, m_{\Phi}\})$, and we have provided the intermediate steps in Appendix \ref{appen2}. The final result can be expressed in terms of the Mandelstam variable
\begin{equation}
    s=-(P_1+P_2)^2 = -(P_3+P_4)^2 \,.
\end{equation}

On contracting the resulting expression of $A_{(0)}^{\alpha\beta\gamma\delta}$ with the polarization vectors $\epsilon^{(0)}_{1,\alpha} \epsilon^{(0)}_{2,\beta} \epsilon^{(0)}_{3,\gamma} \epsilon^{(0)}_{4,\delta}$ we get the desired leading contribution from the Witten diagram that describes the AdS S-matrix as in \ref{defSmatrix}, with the result

\begin{equation}\label{ed.fin}
    \begin{split}
        \mathcal{A}_{(0)}(P_i)&=(g'^2 \phi_0)^2\Big[(\epsilon_{1}^{(0)}.\epsilon_{2}^{(0)})(\epsilon_{3}^{(0)}.\epsilon_{4}^{(0)})\Big] \Bigg[ -\frac{1}{s-m_{\phi}^2}+\frac{1}{R^2}\Bigg( \frac{2 m_{\phi}^4 \left(4 M^2 - m_{\phi}^2\right)}{M^2 (s-m_{\phi}^2)^4}\\
        &+\frac{\big((d-6)m_{\phi}^4 - 2(d-8)m_{\phi}^2 M^2\big)}{M^2(s-m_{\phi}^2)^3}+ \frac{(d-2)(5m_{\phi}^2-6M^2)}{2M^2 (s-m_{\phi}^2)^2}+\frac{(3d-2)}{2 M^2 (s-m_{\phi}^2)}\Bigg)\Bigg]\\
         &-\frac{1}{R^2}\Big[ \big( \epsilon_{3}^{(0)}.\epsilon_{4}^{(0)}\big)\big(\epsilon_{1}^{(0)}.P_2\big)\big(\epsilon_{2}^{(0)}.P_1 \big)\Big]\Bigg[\frac{8}{(s-m_{\phi}^2)^3}+ \frac{4}{M^2(s-m_{\phi}^2)^2}\Bigg].
    \end{split}
\end{equation}

One recognizes that the leading order term in the large $R$ limit gives the expected flat spacetime S-matrix contribution. The subleading terms encode additional contributions about the flat limit due to the AdS potential. \\
\section{The massless limit and the double scaling limit of propagator}\label{limits}

The results in the previous sections pertain to a spin 1 massive vector field with mass $M$, which is related to the conformal dimension $\Delta$ of a dual boundary operator as,
\begin{equation}\label{casimir}
    M^2R^2 =\Delta(\Delta - d) + (d-1)
\end{equation}
In general, primary  operators with conformal dimensions that grow linearly with $R$ correspond to massive particles in the flat patch while considering the large $R$ limit. For massless bulk fields in the flat spacetime limit, we would require operators with finite conformal dimensions. Taking $M=0$ in  \ref{casimir}, one can obtain either  $\Delta = 1 \,\text{or}\, d-1$.

However, the massless limit can't be considered in the AdS S-matrix framework for a few reasons. While we can take $\Delta\rightarrow 1 \,\text{or}\, d-1 $ for the spin 1 vector field, we can't consider $|\mathcal{P}|\rightarrow 0$ as this would make \ref{orm} ill defined and non-invertible. Additionally, with regards to the orientation, the derivation of propagators from the embedding formalism will always lead to timelike particles with $\hat{\mathcal{P}}^2 = 1$. This is at odds with the requirement of null orientations satisfying $\hat{\mathcal{P}}^2 = 0$. One might try to achieve this in the large $R$ limit with an appropriate rescaling $\hat{\mathcal{P}}$ to $\hat{\mathcal{P}}/R$ at the onset, but this would lead to divergences in leading corrections to the propagators about the flat spacetime limit in \ref{VectBulkToBdry} (that go like $M^{-2}$). Thus, although one can derive the AdS S-matrix about a flat limit in the embedding formalism, the prescription of \cite{Gadde:2022ghy} is not suitable for massless fields, as also noted by the authors. 

Alternatively, one could carry out the AdS S-matrix formulation with a different momentum parametrization that has a smooth massless limit, in contrast to \cite{Gadde:2022ghy}. We can parameterize the on-shell timelike momenta in terms of two null vectors as \cite{Kapec:2022axw}
\begin{align}
    {\mathcal{P}}^{\mu}=\omega \tilde{{\mathcal{P}}}^{\mu}=\omega\Big(\hat{q}^{\mu}+\frac{M^2}{\omega^2}n^{\mu}\Big)
\end{align}
with $\hat{q}^{\mu}\cdot n_{\mu}\neq 0$. Note that here the orientation vector $\tilde{{\mathcal{P}}}^{\mu}$ is not the unit normal vector and is related to $\hat{{\mathcal{P}}}^{\mu}$ by $\tilde{{\mathcal{P}}}^{\mu}=\frac{iM}{\omega}\hat{{\mathcal{P}}}^{\mu}$. The detailed analysis of a massless particle scattering by utilizing this momentum parametrization within $1/R$ perturbation theory is beyond the scope of this paper we leave the same for future work.

While a direct $M\rightarrow 0$ limit is not possible with the momentum parametrization of \cite{Gadde:2022ghy}, it might be interesting to consider a double scaling approach.  If we take $R \rightarrow \infty$ and $M\rightarrow 0$, keeping the product $MR=\gamma$ as a large fixed number, while also retaining the orientation $\hat P^2 = 1$ fixed, we get from \ref{fullbdyP}
\begin{equation}
\label{sm}
    \widetilde{\Pi}^{\nu \beta}= e^{i P.X}\left[f_1^{\nu \beta} +{d\over \gamma^2}\hat{P}^{\beta} \hat{P}^{\nu}\right]. 
\end{equation}
One can interpret the second term as a small correction to flat space external leg of a scattering process, originating from the double scaling limit. The scaling dimension of the boundary field will now satisfy
\begin{align}
    \gamma^2 =\Delta(\Delta - d) + (d-1)\,,
\end{align}
which implies $\Delta$ has two solutions. By imposing unitarity we get,
\begin{align}
    \Delta=\frac{d}{2}+\frac{1}{2}\sqrt{(d-2)^2+4\gamma^2}\,.
\end{align}
In $R=1$ units, the above scaling dimension would be that of a massive vector field of mass $\gamma$. However, we note that the double scaling limit effectively reduces a bulk-to-boundary propagator with an arbitrary mass to an external leg with leading small mass contributions in the flat patch, as can be checked by comparing \ref{sm} with \ref{fullbdyP}. In contrast to the massive vector field, the scalar bulk-to-boundary propagator of a massive scalar field given in \cite{Gadde:2022ghy} does not get such a correction in the double scaling limit.
\section{Conclusion}\label{conc}

Building on the formalism of \cite{Gadde:2022ghy}, we have derived an AdS S-matrix for scattering of massive vector fields on AdS spacetimes. This formalism \cite{Gadde:2022ghy} uses Witten diagrams and a conformally covariant momentum space representation for CFT correlators to determine scattering processes about the flat space limit of the bulk AdS spacetime. The ``AdS S-matrix'' can be expanded about the flat patch to derive subleading corrections in powers of the inverse AdS radius. This construction is also wholly consistent with boundary momentum correlation functions and their Witten diagram representation which would typically involve derivatives of the momentum conserving delta functions as corrections about the flat patch. These terms can be derived by using the AdS S-matrix in the Ward identity of the operator $M_{\mu,d+1} = R P_{\mu}$, with this operator arising from the contraction of the conformal group to the Poincare group about the flat patch. \\

We considered the AdS S-matrix formalism to derive massive vector field propagators about the flat space limit, which involve keeping the masses of the particles fixed while taking the AdS radius $R \to \infty$ limit. We specifically derived the bulk-to-boundary propagator and the bulk-to-bulk propagators for the massive vector fields perturbatively up to the sub-leading order in large $R$. Using these propagators, we computed the four-point Witten diagram of the massive spin-$1$ fields, utilizing the action for the Abelian Higgs model on the AdS background. The results of our findings are consistent with the analysis for massive scalar fields in \cite{Gadde:2022ghy}. The Abelian Higgs model involves no cubic or higher self couplings of the vector field, thereby providing no contact or vector exchange diagrams in the four vector AdS diagram. We thus computed the tree-level scattering amplitude resulting from the scattering of four vector fields due to a scalar exchange.\\

Our analysis opens further directions that would be interesting to explore using the AdS S-matrix formalism. This includes the double scaling property for vector fields as discussed in section \ref{limits}. We noted that propagators in this limit appear to be dominated by the leading low mass contribution, and are absent in the massive scalar case of \cite{Gadde:2022ghy}. It would thus be interesting to understand the significance of this limit in vector and more generally non-scalar external fields. \\

Another future direction to consider is the generalization to massless external states. One way in which this could be considered using the current AdS S-matrix formalism is through a parametrization for the massive external states with a smooth massless limit, such as that considered in~\cite{Kapec:2022axw}. Massless external states would also be needed for considering properties of soft factorization on AdS spacetimes.  As we know that
one of the properties of the flat space scattering amplitudes involving massless particles in general are the soft theorems. In its simplest form, the theorem relates the tree level $m+n$ point amplitude of $m-$hard and $n-$soft particles to an universal soft factor and a tree level $m$ point amplitude. The soft factor only depends on the momenta and polarizations of the soft particles and is blind to the details of the interactions involved in the process. The classical limit of the soft theorems were also studied in \cite{Laddha:2017ygw,Laddha:2018rle, Laddha:2018myi, Laddha:2019yaj, Saha:2019tub, Fernandes:2020tsq,Sahoo:2020ryf} and it was shown that the same soft factor can also be obtained from the classical radiation of the fields. A similar analysis has been extended for fields in AdS space in \cite{Banerjee:2020dww, Banerjee:2021llh}.\\

In general, the consecutive flat limit and soft limit do not commute \cite{Chowdhury:2024wwe}. However, about the flat limit, we can consider a simultaneous double scaling limit of taking energy $\omega \to 0$ and the AdS radius $R \rightarrow \infty$, while keeping their product $\omega R = \gamma$ fixed and large. This results in a soft factorization of the $S$-matrix in the flat patch while the flat spacetime soft factors get corrections in inverse powers of $\gamma$ \cite{Banerjee:2020dww, Banerjee:2021llh}. It would thus be interesting to understand the derivation of these factors using appropriate massless extensions of the AdS S-matrix. It would also be valuable to relate derivations from AdS S-matrix with other approaches that utilize Witten diagrams with massless external fields such as \cite{Li:2021snj}, \cite{Chowdhury:2024wwe}, as well as approaches that utilize properties of CFT correlation functions \cite{Marotta:2024sce,Banados:2024kza}. We look forward to exploring these topics in future work.

\section*{Acknowledgement}
 The work of NB is supported by SERB POWER fellowship and we thank people of India for their generous support to the advancement of basic sciences. The work of KF is supported by Taiwan's NSTC with grant numbers 111-2811-M-003-005 and 112-2811-M-003-003-MY3. The work of A. M. is supported by POSTECH BK21 postdoctoral fellowship. A.M. acknowledges support by the National Research Foundation of Korea (NRF) grant funded by the Korean government (MSIT) (RS-2024-00337134 and No. 2022R1A2C1003182). TR acknowledges the support of IISER Bhopal, where the major part of this work was carried out, under the CSIR-UGC grant funded by the Government of India. TR also expresses gratitude to the Junior Research Group (JRG) program at APCTP, Pohang, supported by the Government of South Korea, for their hospitality during the course of this research.
\begin{appendix}

\section{Details of bulk-to-boundary propagator computation}\label{appen1}
To solve the bulk-to-boundary propagator
\begin{align}\label{subleadingsec}
& \eta_{\mu\nu}\Big(-2 M(\hat{P}\cdot\partial)+\partial^2_X\Big) H^{\nu\beta}-\left(\partial_{\nu} \partial_{\mu}+M^{2} \hat{P}_{\mu} \hat{P}_{\nu}-M\left(\hat{P}_{\mu} \partial_{\nu}+\hat{P}_{\nu} \partial_{\mu}\right)\right)H^{\nu\beta}\notag\\
& +\left(d-(d+1)M(X\cdot\hat{P})+M^{2}(X\cdot\hat{P})^{2}\right)\eta_{\mu\nu} G^{\nu\beta}-M X^{\beta} \hat{P}_{\mu}(d+1)+M(X\cdot \hat{P}) \hat{P}^{\beta} \hat{P}_{\mu}=0
\end{align}
 we propose an ansatz
\begin{align}
    H^{\nu\beta}=\alpha(X \cdot\hat{P}) G^{\nu\beta}+\beta X^{\nu}\hat{P}^{\beta}+\gamma X^{\beta}\hat{P}^{\nu}+\sigma \hat{P}^{\nu}\hat{P}^{\beta}\,,
    \label{gbb.ans}
\end{align}
where $G^{\nu\beta}$ is as defined in \ref{G.def},  $\alpha(X \cdot\hat{P})$ is a function of $X.\hat{P}$, while $\beta\,,\gamma $ and $\sigma$ 
 are constants. To simplify the notation in this Appendix, we suppress the explicit dependence on $P$ and $P,X$ in $G^{\nu \beta}$ and $H^{\nu \beta}$ respectively. The derivative terms appearing in the first line of \ref{subleadingsec}  then have the results
\begin{align}
I_1 &= \eta_{\mu\nu}\Big(-2 M(\hat{P}\cdot\partial)+\partial^2_X\Big) H^{\nu\beta}\notag\\&=  \eta_{\mu\nu}G^{\nu\beta}\Big(-2 M(\hat{P}\cdot\partial)+\partial^2_X\Big)\alpha(X \cdot\hat{P}) - 2M\hat{P}^{\beta} \hat{P}_{\mu}(\beta+\gamma)\notag\\&= \eta_{\mu\nu}G^{\nu\beta}\Big(-2 M(\hat{P}\cdot\partial)+\partial^2_X\Big)\alpha(X \cdot\hat{P}) - 2M\hat{P}^{\beta} \hat{P}_{\mu}(\beta+\gamma)
\label{I1.def}
\end{align}
\begin{align}
    I_2 &=-\left(\partial_{\nu} \partial_{\mu}-M\left(\hat{P}_{\mu} \partial_{\nu}+\hat{P}_{\nu} \partial_{\mu}\right)\right)H^{\nu\beta}\notag\\&=-G^{\nu\beta}\left(\partial_{\nu} \partial_{\mu}-M\left(\hat{P}_{\mu} \partial_{\nu}+\hat{P}_{\nu} \partial_{\mu}\right)\right)\alpha(X \cdot\hat{P}) + (d+2) M\beta \hat{P}^{\beta} \hat{P}_{\mu}+\gamma M(\hat{P}^{\beta} \hat{P}_{\mu}+\delta^{\beta}_{\mu})\notag\\&=(d+2) M\beta \hat{P}^{\beta} \hat{P}_{\mu}+\gamma M(\hat{P}^{\beta} \hat{P}_{\mu}+\delta^{\beta}_{\mu})  \label{I2.def}
\end{align}
Note that $G^{\nu\beta}\partial_{\nu} \partial_{\mu}\alpha(X \cdot\hat{P})$ will not contribute since we will get terms involving $G^{\nu\beta}\hat{P}_{\nu}$ which vanishes. On substituting \ref{I1.def} and \ref{I2.def} in \ref{subleadingsec}, and expressing $\delta^{\beta}_{\mu} = \eta_{\mu \nu} G^{\nu \beta} + \hat{P}^{\beta} \hat{P}_{\mu}$,we find the equation

\begin{align}
\eta_{\mu\nu}G^{\nu\beta}&\left[\left(-2 M(\hat{P}\cdot\partial)+\partial^2_X\right) \alpha (X \cdot\hat{P}) + d -(d+1)M(X\cdot\hat{P})+M^{2}(X\cdot\hat{P})^{2} + \gamma M\right] \notag\\
& - X^{\beta}\hat{P}_{\mu} \left(M^2 \gamma + (d+1) M\right)  - (X.\hat{P}) \hat{P}^{\beta} \hat{P}_{\mu} \left(M^2 \beta - M\right) +  \hat{P}^{\beta} \hat{P}_{\mu} \left( -\sigma M^2 + d M\right) = 0 \,.
\label{gbb.me}
\end{align}

The desired solution follows from setting the coefficients of the tensor blocks to vanish. The coefficients of $X_{\mu}\hat{P}^{\beta},  (X\cdot\hat{P}) \hat{P}^{\beta}\hat{P}_{\mu}$ and $\hat{P}^{\beta}\hat{P}_{\mu}$ are readily solved to give

\begin{equation}
\gamma=-\frac{1}{M}(d+1) \,, \quad \beta = \frac{1}{M}\,, \quad \sigma = \frac{d}{M^2}
\label{bgs.sol}
\end{equation}

Substituting $\gamma\,, \beta$ and $\sigma$ in the coefficient for $\eta_{\mu\nu}G^{\nu\beta}$ in \ref{gbb.me} then gives a differential equation

\begin{equation}
\left(-2 M(\hat{P}\cdot\partial)+\partial^2_X\right) \alpha (X \cdot\hat{P}) - (d+1)M(X\cdot\hat{P})+M^{2}(X\cdot\hat{P})^{2} - 1 = 0 \,.
\label{alph.de}
\end{equation}

Requiring $\alpha(0) = 0$, the solution of \ref{alph.de} is

\begin{equation}
\alpha(X \cdot\hat{P}) = -\frac{(d+2) (\hat{P}\cdot X)}{4M} -\frac{d (\hat{P}\cdot X)^2}{4} + \frac{M (\hat{P}\cdot X)^3 }{6}
\label{alpha.sol}
\end{equation}

On substituting \ref{bgs.sol} and \ref{alpha.sol} in \ref{gbb.ans}, we have the $1/R^2$ correction of the bulk-to-boundary propagator

\begin{equation}
    \begin{split}
        H^{\nu \beta} &= \left[-\frac{(d+2) (\hat{P}\cdot X)}{4M} -\frac{d (\hat{P}\cdot X)^2}{4} + \frac{M (\hat{P}\cdot X)^3 }{6}\right] G^{\nu \beta} \\
&\qquad \quad  - \frac{(d+1)}{M } X^{\beta} \hat{P}^{\nu} + \frac{1}{M} X^{\nu} \hat{P}^{\beta} + \frac{d}{M^2} \hat{P}^{\beta} \hat{P}^{\nu}
    \end{split}
\end{equation}
\section{Evaluation of \ref{ed.fin}}\label{appen2}
We can simplify each of the terms appearing inside the parenthesis of \ref{abhiggs2}. For the vertex operator for the incoming particles, we have from \ref{v12.exp} the following expression
\begin{align}
&\sqrt{-g(-i\partial_{\mathfrak{p}_1})} V_{12}^{\alpha\beta}(-i\partial_{\sigma_1},-i\partial_{\sigma_2},-i\partial_{\mathfrak{p}_1}) =  G^{\mu \alpha}(P_1) G^{\nu \beta}(P_2) \eta_{\mu \nu}\left(1 - \frac{\partial_{\mathfrak{p}_1^2}}{2 R^2}\right)\notag\\& + \frac{1}{R^2}\left(\partial_{\mathfrak{p}_1}^{\alpha}\partial_{\mathfrak{p}_2}^{\beta}+{1\over M^2}{P_1}^{\alpha}\partial_{\mathfrak{p}_1}^{\beta}\partial_{\sigma_1}+{1\over M^2}P_2^{\beta}\partial_{\mathfrak{p}_1}^{\alpha}\partial_{\sigma_2}+{1\over M^4}{P_1}^{\alpha}P_2^{\beta}\partial_{\sigma_1}\partial_{\sigma_2}\right)\notag\\
& \; \; + \eta_{\mu \nu}\frac{G^{\mu \alpha}(P_1) G^{\nu \beta}(P_2)}{M^2 R^2}\left(\frac{(d+2)}{4} \left(\partial_{\sigma_1} + \partial_{\sigma_2}\right) -\frac{d}{4}\left(\partial_{\sigma_1}^2 + \partial_{\sigma_2}^2\right) - \frac{1}{6}\left(\partial_{\sigma_1}^3 + \partial_{\sigma_2}^3\right)\right) \notag\\
&\quad + \eta_{\mu \nu}\frac{G^{\nu \beta}(P_2)}{M^2 R^2} \left((d+1) {P_1}^{\mu} \partial_{\mathfrak{p}_1}^{\alpha}-{P_1}^{\alpha}\partial_{\mathfrak{p}_1}^{\mu}  - \frac{d}{M^2} {P_1}^{\alpha} {P_1}^{\mu}\right) \notag\\
& \qquad + \eta_{\mu \nu}\frac{G^{\mu \alpha}(P_1)}{M^2 R^2} \left((d+1) {P_2}^{\nu}\partial_{\mathfrak{p}_1}^{\beta} - {P_2}^{\beta}\partial_{\mathfrak{p}_1}^{\nu}  - \frac{d}{M^2} {P_2}^{\beta} {P_2}^{\nu}\right) 
\label{v12.expder}
\end{align}
On using the expression for $G^{\mu \nu}$ from \ref{G.def}, we then find
\begin{align}
&\sqrt{-g(-i\partial_{\mathfrak{p}_1})} V_{12}^{\alpha\beta}(-i\partial_{\sigma_1},-i\partial_{\sigma_2},-i\partial_{\mathfrak{p}_1}) \notag\\&=  \left(\eta^{\alpha\beta}+\frac{1}{M^2}({P_1}^{\alpha}{P_1}^{\beta}+{P_2}^{\alpha}{P_2}^{\beta})+\frac{1}{M^4}(P_1\cdot P_2){P_1}^{\alpha}{P_2}^{\beta}\right) \left[1 - \frac{\partial_{\mathfrak{p}_1^2}}{2 R^2}\right.\notag\\&\left.+\frac{1}{M^2 R^2}\left(\frac{(d+2)}{4} \left(\partial_{\sigma_1} + \partial_{\sigma_2}\right) -\frac{d}{4}\left(\partial_{\sigma_1}^2 + \partial_{\sigma_2}^2\right) - \frac{1}{6}\left(\partial_{\sigma_1}^3 + \partial_{\sigma_2}^3\right)\right)\right]\notag\\&+ \frac{1}{R^2}\left(\partial_{\mathfrak{p}_1}^{\alpha}\partial_{\mathfrak{p}_2}^{\beta}+{1\over M^2}{P_1}^{\alpha}\partial_{\mathfrak{p}_1}^{\beta}\partial_{\sigma_1}+{1\over M^2}P_2^{\beta}\partial_{\mathfrak{p}_1}^{\alpha}\partial_{\sigma_2}+{1\over M^4}{P_1}^{\alpha}P_2^{\beta}\partial_{\sigma_1}\partial_{\sigma_2}\right)\notag\\
& \quad + \frac{1}{M^2 R^2} \left((d+1) ({P_1}^{\beta} \partial_{\mathfrak{p}_1}^{\alpha}+{P_2}^{\alpha} \partial_{\mathfrak{p}_1}^{\beta})-({P_1}^{\alpha}\partial_{\mathfrak{p}_1}^{\beta}+{P_2}^{\beta}\partial_{\mathfrak{p}_1}^{\alpha})  - \frac{d}{M^2} ({P_1}^{\alpha} {P_1}^{\beta}+{P_2}^{\alpha} {P_2}^{\beta})\right. \notag\\
&\left. + \frac{(d+1)}{M^2} (P_1\cdot P_2)({P_2}^{\beta}\partial_{\mathfrak{p}_1}^{\alpha}+{P_1}^{\alpha}\partial_{\mathfrak{p}_1}^{\beta})- \frac{{P_1}^{\alpha}{P_2}^{\beta}}{M^2}({P_2}\cdot\partial_{\mathfrak{p}_1}+{P_1}\cdot\partial_{\mathfrak{p}_1}) - \frac{2d}{M^4}(P_1\cdot P_2) {P_2}^{\beta} {P_1}^{\alpha}\right)   \label{v12.expder1}
\end{align}
In a similar manner the other vertex contribution in \ref{abhiggs2} contains no derivatives and results in 
 \begin{equation}\label{v34}
    \begin{split}
        V_{34}^{\gamma\delta}(0,0,0)&=G^{\gamma\rho}(P_3)G^{\delta\sigma}(P_4)\eta_{\rho\sigma}-{1\over R^2}\Big\{ \frac{d}{M^4}\left({P}_3^{\gamma}{P}_3^{\delta}+{P}_4^{\delta}{P}_4^{\gamma}\right)+ \frac{2d}{M^6}({P}_3\cdot {P}_4){P}_3^{\gamma}{P}_4^{\delta}\Big\}\\&=\left(\eta^{\gamma\delta}+\frac{1}{M^2}({P_3}^{\gamma}{P_3}^{\delta}+{P_4}^{\gamma}{P_4}^{\delta})+\frac{1}{M^4}(P_3\cdot P_4){P_3}^{\gamma}{P_4}^{\delta}\right)\\&-{1\over R^2}\Big\{ \frac{d}{M^4}\left({P}_3^{\gamma}{P}_3^{\delta}+{P}_4^{\delta}{P}_4^{\gamma}\right)+ \frac{2d}{M^6}({P}_3\cdot {P}_4){P}_3^{\gamma}{P}_4^{\delta}\Big\}
    \end{split}
\end{equation}
\par
The last piece in \ref{abhiggs2} involving the bulk-to-boundary propagator can be further reduced by utilizing its functional dependence only on $\alpha$,
\begin{equation}\label{opgbb}
    \begin{split}
    &{g}_{\rm BB}(0,0,\partial_{\alpha})\frac{1}{\Pi^2+\alpha m_{\Phi}^2}\\&=\frac{1}{\Pi^2+\alpha m_{\Phi}^2}+{1\over R^2}\left[-\frac{(d-1)\Pi^2}{(\Pi^2+\alpha m_{\Phi}^2)^3}+\frac{(d-7)m_{\Phi}^2\Pi^2+(d+1)m_{\Phi}^4}{(\Pi^2+\alpha m_{\Phi}^2)^4}\right]
\end{split}
\end{equation}
The main advantage of following the AdS S-matrix formulation is that we can express the amplitude in \ref{amp} as a function of Mandelstam variables. For the scalar exchange diagram we only need to introduce
\begin{equation}
    s=-(P_1+P_2)^2 = -(P_3+P_4)^2
\end{equation}
To evaluate the action of derivatives given in \ref{v12.expder1} we consider some useful relations. From the expression of $\Pi^{\mu} = \sigma_1 P_1^{\mu}+ \sigma_2 P_2^{\mu} + \mathfrak{p}_i^{\mu}$ , we find
\begin{equation}
    \begin{split}
        \Pi^2=-M^2(\sigma_2-\sigma_1)^2-s\sigma_1\sigma_2+2\mathfrak{p}_1\cdot \left(\sigma_1 P_1+\sigma_2 P_2 + \frac{\mathfrak{p}_1}{2}\right)\,,
    \end{split}
\end{equation}
where we used $2P_1\cdot P_2=-(s-2M^2)$ and $ P_1^2 = -M^2 = P_2^2$. We then have five kinds of derivatives that will be useful in the derivation, that either act on $\Pi^{\mu}$ and $\Pi^2$. These are
\begin{align}
\partial_{\mathfrak{p}_1}^{\alpha}\Pi^{2}=2\Pi^{\alpha}\,, &\qquad 
\partial_{\mathfrak{p}_1}^{\alpha}\Pi^{\mu}=\eta^{\mu\alpha}\,, \qquad \partial_{\sigma_1}\Pi^{2}\Big\vert_L=\partial_{\sigma_2}\Pi^{2}\Big\vert_L=-s \notag\\
\partial_{\sigma_1} \partial_{\sigma_2}\Pi^{2}\Big\vert_L & = -s + 2 M^2 \,, \qquad \partial_{\sigma_1}^2 \Pi^2\Big\vert_L = -2 M^2 = \partial_{\sigma_2}^2 \Pi^2\Big\vert_L
\end{align}

Taking the product of \ref{v12.expder}, \ref{v34} and \ref{opgbb}, and retaining terms up to $R^{-2}$ order, we then get the expression for $A^{\alpha \beta \gamma \delta}_{(0)}$. Contracting this with the polarizations for the external fields then provides the AdS S-matrix for the process as given in \ref{ed.fin}.

\end{appendix}
\bibliography{ads.bib}

\end{document}